%% file: acl2026.tex
\documentclass[11pt]{article}

\usepackage[preprint]{acl}

\usepackage{times}
\usepackage{latexsym}
\usepackage{algorithm}
\usepackage{booktabs}   %% For formal tables:
\usepackage{subcaption} %% For complex figures with subfigures/subcaptions
\usepackage{array}
\usepackage{amsmath,amsfonts}
\usepackage{algpseudocode}
\usepackage{graphicx}
\usepackage{threeparttable}
\usepackage{textcomp}
\usepackage{float}
\usepackage{listings}
\usepackage{xspace}
\usepackage{multirow}
\usepackage{amsthm}

\usepackage{amssymb} 
\usepackage{balance}
\usepackage{pgfplots}
\usepackage{pgfplotstable}
\usepackage{tabularx}
\usepackage{wrapfig}
\usepackage{comment}
\usepackage[T1]{fontenc}
\usepackage[utf8]{inputenc}

\usepackage{microtype}

\usepackage{inconsolata}

\usepackage{graphicx}
\newcommand{\tool}{HxAgent\xspace}

\definecolor{lightgreen}{rgb}{0.56, 0.93, 0.56}
\definecolor{lightcoral}{rgb}{0.94, 0.5, 0.5}
\definecolor{lightercoral}{rgb}{0.98, 0.7, 0.7}
\definecolor{lightblue}{rgb}{0.68, 0.85, 0.90}
\definecolor{darkerlightblue}{rgb}{0.55, 0.75, 0.85}
\definecolor{codegray}{gray}{0.5}
\definecolor{codepurple}{rgb}{0.58,0,0.82}
\definecolor{backcolour}{rgb}{0.95,0.95,0.92}
\newcommand{\bluetext}[1]{\textcolor{blue}{#1}}

\lstdefinestyle{htmlStyle}{
    language=HTML,
    numbers=left,
    numberstyle= \tiny,
    keywordstyle= \color{blue!70},
    commentstyle= \color{red!50!green!50!blue!50},
    stringstyle=\color{red!70},
    frame=shadowbox,
    rulesepcolor= \color{red!20!green!20!blue!20} ,
    xleftmargin=1.5em,xrightmargin=0em, aboveskip=1em,
    framexleftmargin=1.5em,
            numbersep= 5pt,
    basicstyle=\scriptsize\ttfamily,
    numberstyle=\scriptsize\ttfamily,
    emphstyle=\bfseries,
    morekeywords={doctype, html, head, body, title, script, div, span, a, data-result} % Additional HTML keywords
}

\lstdefinestyle{textStyle}{
    keywordstyle= \color{blue!70},
    commentstyle= \color{red!50!green!50!blue!50},
    stringstyle=\color{red!70},
    frame=shadowbox,
    rulesepcolor= \color{red!20!green!20!blue!20} ,
    xleftmargin=1.5em,xrightmargin=0em, aboveskip=1em,
    framexleftmargin=1.5em,
            numbersep= 5pt,
    basicstyle=\scriptsize\ttfamily,
    numberstyle=\scriptsize\ttfamily,
    emphstyle=\bfseries,
    numbers=none,
    breaklines=true,                  % Turn on breaking
    breakatwhitespace=false,          % Force break even if there is no space
    columns=fullflexible,             % Allow character compression
    linewidth=\columnwidth,           % Strictly limit width to the column
}

\definecolor{custom-blue}{rgb}{0,0,0}
\definecolor{my-blue}{rgb}{0,0,0}
\newcommand{\code}[1]{{\footnotesize\texttt{#1}}}
\lstdefinestyle{pythonStyle}{
    backgroundcolor=\color{backcolour},
    commentstyle=\color{codegray},
    keywordstyle=\color{magenta},
    numberstyle=\tiny\color{codegray},
    stringstyle=\color{codepurple},
    basicstyle=\ttfamily\scriptsize ,
    breakatwhitespace=false,
    breaklines=true,
    captionpos=b,
    keepspaces=true,
    numbers=left,
    numbersep=5pt,
    showspaces=false,
    showstringspaces=false,
    showtabs=false,
    tabsize=2,
    language=Python,
    escapeinside={(*@}{@*)}
}

\usepackage{xcolor}

\definecolor{eclipseStrings}{RGB}{42,0.0,255}
\definecolor{eclipseKeywords}{RGB}{127,0,85}
\colorlet{numb}{magenta!60!black}

\lstdefinelanguage{json}{
    basicstyle=\scriptsize\ttfamily,
    numberstyle=\scriptsize\ttfamily,
    emphstyle=\bfseries,
    commentstyle=\color{eclipseStrings}, % style of comment
    stringstyle=\color{eclipseKeywords}, % style of strings
    numbers=left,
    numberstyle= \tiny,
    keywordstyle= \color{blue!70},
    commentstyle= \color{red!50!green!50!blue!50},
    frame=shadowbox,
    rulesepcolor= \color{red!20!green!20!blue!20} ,
    xleftmargin=1.5em,xrightmargin=0em, aboveskip=1em,
    framexleftmargin=1.5em,
            numbersep= 5pt,
    showstringspaces=false,
    breaklines=true,
    string=[s]{"}{"},
    comment=[l]{:\ "},
    morecomment=[l]{:"},
    literate=
        *{0}{{{\color{numb}0}}}{1}
         {1}{{{\color{numb}1}}}{1}
         {2}{{{\color{numb}2}}}{1}
         {3}{{{\color{numb}3}}}{1}
         {4}{{{\color{numb}4}}}{1}
         {5}{{{\color{numb}5}}}{1}
         {6}{{{\color{numb}6}}}{1}
         {7}{{{\color{numb}7}}}{1}
         {8}{{{\color{numb}8}}}{1}
         {9}{{{\color{numb}9}}}{1}
}

\usepackage{tikz}

\newcommand{\circlednum}[2][red]{%
\tikz[baseline=(char.base)]{
    \node[shape=circle, fill=#1, text=white, inner sep=1pt] (char) {#2};}%
}

\definecolor{request}{HTML}{D80073}
\definecolor{response}{HTML}{0050EF}
\title{HxAgent: Iterative Agent Planning for End-to-End\\ Web Application Testing}

\author{
 \textbf{Tu Nguyen\textsuperscript{1,2}},
 \textbf{Duy Cao\textsuperscript{1,2}},
 \textbf{Viet Nguyen\textsuperscript{1,2}},
 \textbf{Phu Nguyen\textsuperscript{1,2}},
 \\
 \textbf{Vy Le\textsuperscript{2,3}},
 \textbf{Nguyen TK Nguyen\textsuperscript{1,2}},
 \textbf{Tien N. Nguyen \textsuperscript{4}},
 \textbf{Vu Nguyen\textsuperscript{1,2}}
\\
\\
 \textsuperscript{1}University of Science, Ho Chi Minh city, Vietnam,
 \\
 \textsuperscript{2}Vietnam National University, Ho Chi Minh city, Vietnam,
 \\
 \textsuperscript{3}University of Information Technology, Ho Chi Minh city, Vietnam,
 \\
 \textsuperscript{4}University of Texas at Dallas, USA,
\\
 \small{
   \textbf{Correspondence:} 
   \href{mailto:nvu@fit.hcmus.edu.vn}
   {Vu Nguyen (nvu@fit.hcmus.edu.vn)}
 }
}

\begin{document}
\maketitle

\begin{abstract}
\input{sections/abstract}
\end{abstract}

\section{Introduction}
\input{sections/intro}

\input{sections/key-ideas}

\input{sections/overview}

\input{sections/experiment}

\section{Empirical Results}
\input{sections/rq1}
\input{sections/rq2}
\input{sections/rq-test}

\input{sections/rq3}
\input{sections/rq4}

\input{sections/rq5}
\input{sections/rq6}
%\label{sec:threat}
%\input{sections/threats-to-validity}

\section{Related Work}
\label{sec:related_work}
\input{sections/related}
\section{Conclusion}
\label{sec:conclusion}
\input{sections/conclusion_future_work}

\input{sections/threats-to-validity}

\newpage
% \balance
%\bibliographystyle{ACM-Reference-Format}
\bibliography{references,references-visioflow,references-tessara}
\newpage

\appendix
\input{sections/motiv-icse25}
\input{sections/experience-memory}

\input{sections/planning_appendix}

\input{sections/dataset_appendix}

\input{sections/results_analyst}

% \section{Example Appendix}
% \label{sec:appendix}

\end{document}

%% file: sections/abstract.tex
In automated web testing, generating test cases and performing testing using  functionality descriptions in natural-language is crucial for improving efficacy. These tasks require such a testing agent to carry out tasks on the target application and generating tests autonomously. 
We introduce {\tool}, an iterative LLM-based planning agent with a proactive correction strategy. After each step, {\tool} reassesses the web state to determine the next action using (1) current observations, (2) short-term memory of past actions, and (3) long-term experience extracted from past (in)correct sequences of actions. {\tool} achieves 97.4\% Exact-Match accuracy on MiniWoB++, comparable to the best baselines without human demonstrations and surpassing the recent WALT by 10.5\%. On a dataset of 350 web tasks, it attains 83.8\% Exact-Match and 91.8\% Prefix-Match, exceeding WALT by 13.4\%. On OnlineMind2Web, it further improves over WALT by 4.6\%.

%% file: sections/intro.tex
In web UI testing, much of the process remains manual and error-prone. Testers execute textual task descriptions using tools such as Selenium~\cite{selenium} or Katalon~\cite{katalon}, which record user interactions to form Sequences of Actions (SoAs)—automated test scripts. While effective, this manual recording is time-consuming and susceptible to human error, motivating automation.

Recent LLM-based approaches~\cite{humphreys2022data,jia2018dom,liu2018reinforcement,drouin2024workarena,liu2023agentbench,sumerscognitive} advance automated interaction but cannot directly generate executable test scripts. Existing UI agents face four main limitations:
(1) Models such as WebArena~\cite{zhou2023webarena} and WebLINX~\cite{lu2024weblinx} prioritize task completion over capturing executable SoAs, focusing on end-user automation rather than test generation.
(2) Visual and multimodal planners (e.g., SeeClick~\cite{cheng-etal-2024-seeclick}, UI-TARS~\cite{qin2025uitarspioneeringautomatedgui}, SeeAct~\cite{pmlr-v235-zheng24e}, CogAgent~\cite{hong2024cogagentvisuallanguagemodel}) represent actions as absolute screen coordinates or textual descriptions—fragile under UI changes and difficult to translate into executable scripts.
(3) Pattern-based agents (e.g., MindAct~\cite{deng2024mind2web}, RCI~\cite{kim2024language}, AdaPlanner~\cite{sun2024adaplanner}, Synapse~\cite{zheng2023synapse}) depend heavily on human demonstrations, limiting scalability.
(4) Li {\em et al.}~\cite{li2023zero} avoid demonstrations via self-reflection, yet often fall into unguided trial-and-error loops lacking reasoning over prior state–action pairs.

We introduce {\tool}, an iterative LLM-based planning approach for automatically generating executable SoAs from task goals. Inspired by ReAct~\cite{yao2023react}, {\tool} continuously reasons over (1) the current page, (2) feasible actions, (3) short-term memory of prior state–action pairs, and (4) long-term experience with rule abstractions extracted from past correct and incorrect trajectories. Each action is executed and its outcome informs the next step until completion. By anchoring reasoning in textual and visual evidence, {\tool} mitigates hallucinations and ensures decisions remain grounded in real screen context.

Unlike reactive reflection methods, {\tool} performs proactive correction—reassessing and adjusting its plan after each step. It restricts actions to feasible, executable elements, and combines short-term memory with long-term experience to enhance reasoning and stability.

Empirically, {\tool} outperforms baselines by 19.3\%, 5.7\%, and 11.4\% in \code{Exact-Match} accuracy on OnlineMind2Web, MiniWoB++, and Real-world datasets, respectively, and achieves 77.3\%–100.0\% success in test script generation. On challenging tasks, it attains 81.6\% Exact-Match, while others largely fail.
In brief, this paper provides the following contributions:

{\bf 1. {\tool}: an LLM-based approach} to support automated web testing by generating the sequences of actions to perform on a web application to accomplish a given task.

{\bf 2. An iterative LLM-based agent planning} with both short-term and long-term experience.

{\bf 3. An empirical evaluation} to show its effectiveness over the baselines and {\bf a dataset}~\cite{project-website} containing tasks and actual SoAs, which can serve as a benchmark for future studies.

%% file: sections/key-ideas.tex
\section{Key Ideas}

% A brief intro to why come up to key ideas
We design {\tool} with the following key ideas:

\textbf{Key Idea 1 [{\bf Iterative LLM-based Agent Planning}]}. To model executable action sequences, we design an iterative LLM-based planning mechanism that controls the next action in a feedback loop, grounding the model in textual and visual screen context to avoid irrelevant or hallucinated actions. {\tool} determines each action from four inputs: the current page, feasible actions, memory of previous state–action pairs, and experience from past (in)correct sequences. Each action is executed in a testing environment, and the resulting state guides the next iteration until completion. Unlike prior UI agents, actions are explicitly represented as tuples of operations on web elements with optional input values. Unlike self-reflection approaches, {\tool} performs proactive correction which reassesses and adjusts its plan after each step rather than re-executing faulty plans after delay.

\textbf{Key Idea 2 [{\bf Sequence action memory} as short-term memory]}. A crucial aspect of navigating complex web tasks is maintaining a persistent awareness of previous steps and the evolving states of webpages. Drawing inspiration from this necessity, we introduce a short-term memory mechanism that involves feeding LLMs with information about the previous steps taken and the corresponding states of the website after each action. {\tool} maintains a structured memory of prior interactions, allowing it to avoid previously failed decisions and inform future reasoning with explicit historical context. 

%Li et al.’s method uses a limited memory for "disabled actions" but lacks a broader awareness of prior state-action dynamics, making it prone to cyclical failures.}}

%As in Fig.~\ref{fig:motiv2}, each action and the corresponding webpage after that are recorded in terms of json files and xpath locators as a short-term external memory. 

%By incorporating this trajectory memory component, we expect that the LLM can accumulate knowledge over successive interactions, enabling more informed decision-making.

% A crucial aspect of navigating complex web tasks is maintaining a persistent awareness of previous steps and the evolving states of webpages. Drawing inspiration from this necessity, we introduce a short-term memory mechanism to assist our iterative planning process. This mechanism involves feeding the LLMs with information about the previous steps taken and the corresponding states of the website after each action. 
%By incorporating this trajectory memory component, LLMs can accumulate knowledge over successive interactions, enabling more informed decision-making processes.

%To reinforce the experience for the central LLM in the iterative agent planning process, we record the entire sequences of state-action pairs that have been taken and the corresponding correct/incorrect status. 

\textbf{Key Idea 3 [{\bf Combining action memory with experience reinforcement}]}.
Building on LLM reflection mechanisms~\cite{shinn2024reflexion, li2023zero}, we propose a two-phase strategy with a training and an evaluation phase. In training, the LLM gathers state–action sequences with success or failure feedback, which are aggregated to guide performance in evaluation. {\em Short-term memory} complements this process by retaining recent actions and outcomes, such as clicks and inputs, providing continuity, while accumulated experience enables the model to generalize from past patterns and apply effective strategies in new tasks.

% {\color{custom-blue}{Building upon LLMs' reflection~\cite{shinn2024reflexion} as in~\cite{li2023zero}, we propose a two-phased strategy consisting of a training phase and an evaluation phase. During the training phase, LLMs rely on their reflection capabilities to gather the entire sequences of prior state-action pairs the corresponding correct/incorrect status. The knowledge from this phase is aggregated to enhance performance during the evaluation phase. In practice, the judgment of sequences can be part of the process where testers evaluate them. 

% {\em Short-term memory complements experience to enhance the ability of the LLM to learn from recent interactions and make informed decisions for future actions}. Short-term memory enables the model to retain information about the most recent actions and their outcomes, providing essential context and continuity. For instance, when navigating a series of webpages to complete a task, short-term memory allows the LLM to remember recent clicks, form inputs, and navigation choices. Meanwhile, experience draws on a broader set of patterns and knowledge from past interactions, helping the model recognize similarities with previous tasks and apply effective strategies.}}

%This accumulated experience aids in predicting successful actions based on previous analogous situations.

% TODO: Add the key idea explanation for vision
\textbf{Key Idea 4 [\textbf{Vision enhancement}]}. The HTML of real-world websites can be extensive, risking exceeding the LLM's context length. Selecting elements is also challenging when complex layouts obscure an element's context: in a date picker, for instance, an agent may detect "Day 1" as interactable but fail to distinguish between months, yielding ambiguous options. To address this, we exploit the multi-modal capabilities of LLMs, providing the screenshot alongside the actionable elements of the current webpage to support the choice of next action.

% Building upon the reflection ability of LLMs, we propose a two-phased strategy consisting of a training phase and an evaluation phase. During the training phase, without any expert demonstrations, LLMs rely on their reflection capabilities to gather insights from past attempts. The knowledge accumulated during this phase, whether from failed or successful attempts, is aggregated to enhance performance during the evaluation phase. 
%By leveraging the reflective nature of LLMs, our approach aims to make the system independent from expert guidance.

%=====================================================================================================
%\subsubsection{Key Idea 4 [Representing web state for agent planning]} We recognize the challenge posed by raw HTML representations, which contains much redundancy that might result in hallucination in LLMs. Therefore, we propose a method focused on extracting only the %interactive elements and their context.

% We recognize the challenge posed by raw HTML representations, which contains much redundancy that might result in hallucination in LLMs. Therefore, we propose a method focused on extracting only the interactive elements and their contextual information. By selectively filtering HTML content, we attempt to make all information required to complete the tasks fit into LLMs' context length, reduce distractions, and minimize the occurrence of hallucinations. This approach ensures that the LLM's attention remains concentrated on the most relevant aspects of the web environment, enhancing its ability to make informed decisions.

%% file: sections/overview.tex
\section{\tool}
\label{sec:overview}

\subsection{Formulation}
\begin{figure*}[htpb]
    \centering
    \includegraphics[width=0.98\textwidth]{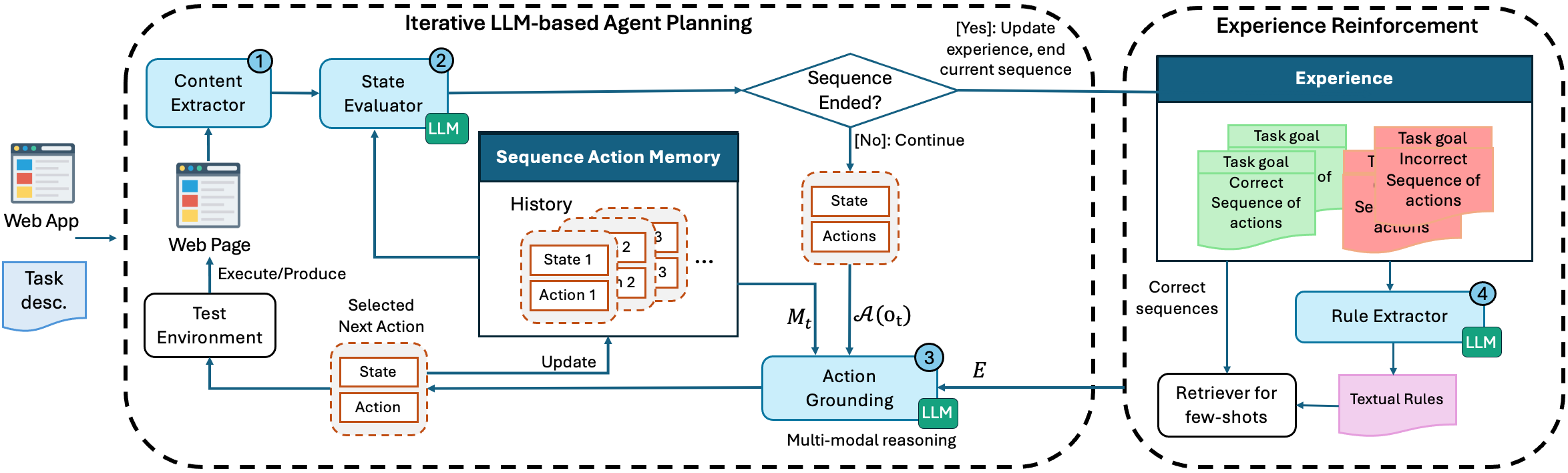} 
    \vspace{-6pt}
    \caption{{\tool} Architecture}
    \vspace{-12pt} 
    \label{fig:overview}
\end{figure*}

We model web-based task execution as a Partially Observable Markov Decision Process (POMDP)
$\mathcal{M} = (\mathcal{S}, \mathcal{A}, \mathcal{O}, \mathcal{T}, \mathcal{R}, \gamma)$
where $s_t \in \mathcal{S}$ denotes the latent web environment state at time $t$, 
$a_t \in \mathcal{A}$ is an executable action, $o_t \in \mathcal{O}$ is the observation, 
$\mathcal{T}$ is the transition function, $\mathcal{R}$ is the reward function, and $\gamma \in (0,1]$ is the discount factor.

\textbf{Multi-modal observation.} Since the agent can only partially observe $s_t$, it receives a multi-modal observation $o_t = \Omega(s_t) = \big(o_t^{\text{txt}},\, o_t^{\text{img}},\, o_t^{\text{elm}}\big)$.
where $o_t^{\text{txt}}$ contains textual DOM information, $o_t^{\text{img}}$ is a webpage screenshot, and
$o_t^{\text{elm}}$ denotes the set of interactable elements.

The feasible action set is defined as
%\begin{equation}
$\mathcal{A}(o_t) = \{(op, e, v) \mid e \in o_t^{\text{elm}},\; op \in \mathcal{O},\; v \in \mathcal{V} \cup \{\varnothing\}\}$
%\end{equation}
where $op$ denotes an operation (e.g., click or input), $e$ a web element, and $v$ an optional input value.

\textbf{Sequence of actions.} An action sequence to perform a task up to time $t$ is defined as
\begin{equation}
\tau_t = \big( (s_0, a_0), (s_1, a_1), \ldots, (s_t, a_t) \big)
\end{equation}
where the state transitions $s_{t+1} \sim \mathcal{T}(s_t, a_t)$ and $a_t$ is determined using the grounding function (\ref{eq_action_grounding}). 

\textbf{Sequence action memory (SAM).} This short-term memory at step $t$ is defined as 
\begin{equation}
M_t = \{(\hat{s}_{t-k}, a_{t-k}), \ldots, (\hat{s}_{t-1}, a_{t-1})\}
\end{equation}
where $\hat{s}_t$ denotes the observed state representation and $k$ is the memory window size.
This memory maintains recent interaction context to guide subsequent decisions to accomplish the current task.

\textbf{Long-term experience memory.} This memory stores historical interaction trajectories from multiple tries for each task, their outcomes, and behavioral rules:
\begin{equation}
E = \{(\tau^{(i)}, y^{(i)})\}_{i=1}^{N} \cup \mathcal{K}^{(j)}
\end{equation}
where $\tau^{(i)}$ denotes a complete SoA and
$y^{(i)} \in \{0,1\}$ indicates task success or failure. Abstracted behavioral rules can be derived as
\begin{equation}
\mathcal{K}^{(j+1)} = \phi(\tau^{(i)}, E)
\end{equation}
where $\phi$ denotes a rule-based knowledge extraction function. {\color{custom-blue}{This function is implemented by prompting the LLM with a completed trajectory and the current rule set, asking it to abstract a concise behavioral rule (e.g., "verify the label before clicking") only when the candidate rule is not already entailed by an existing one. Rules are stored as plain-text heuristics alongside their source SoA. Implementation details are given in Appendix~C.}}

These memory components complement each other: the short-term memory maintains immediate context and the current sequence of actions required to accomplish the task, while the long-term experience captures learned strategies distilled from many SoAs explored during training.

\textbf{Action grounding.} Inspired by ReAct~\cite{yao2023react}, which enables LLMs to interleave reasoning and acting by pausing after each decision to observe the environment, we adopt an iterative action grounding scheme that conditions decision making on the current observation $o_t$, short-term memory $M_t$, and accumulated experience $E$. 
At a time step $t$, the agent selects an action according to
\begin{equation}
\label{eq_action_grounding}
a_t \sim \pi(. \mid o_t, \mathcal{A}(o_t), M_t, E)
\end{equation}
where the grounding function $\pi$ is implemented by an LLM and constrained to the feasible action set $\mathcal{A}(o_t)$. This formulation enables iterative, memory-aware, and experience-driven decision making to support long-horizon task execution in dynamic web environments.

\textbf{State transition.} After grounding an action, the agent executes the action on the test environment and transitions to the next state $s_{t+1} \sim \mathcal{T} (s_t, a_t)$. 

\textbf{State evaluation.}
At each new state, the agent evaluates the state to determine whether the task has been successfully completed or whether execution should terminate early due to failure. The evaluation is defined as follows:
\begin{equation}
\{ \text{success}, \text{stop} \} \sim \mathcal{G}(\tau_t, M_t, E)
\end{equation}
If the task is completed, the outcome $y^{(i)}$ of the experience memory $E$ is updated automatically when human judgment is not involved in determining task success.  

%$\{complete, stop\} \sim \mathcal{g}(\tau_t, M_t, E)$.

\subsection{Architecture} 
Fig.~\ref{fig:overview} shows {\tool}'s architecture with LLM indicating LLM-based agents. It takes as input the web application and the task description, and produces the SoA to complete the task. {\tool} operates with two key components: 1) Iterative LLM-based Agent Planning to produce action sequences, and 2) Experience Reinforcement to support the decision-making.

%From the target Web application and task description, {\tool} initiates the iterative process using the specified starting page URL. The process starts with Content Extractor \circlednum[response]{1} operating on the current webpage. The HTML content of the current page is recorded, along with a set of feasible actions expressed in the JSON format. The process then continues with State Evaluator \circlednum[response]{2} which uses LLMs to analyze the current state of the action sequence. In one prompt, the agent performs the following: summarize previous interactions for the task, describe the current state, determine whether to stop the task early due to significant divergence from the expected state, decide whether the task is completed, and determine if the task is successfully achieved.

Given a target web application and task description, {\tool} initiates its iterative process from the specified starting page URL. The process begins with Content Extractor \circlednum[response]{1}, which operates on the current webpage state $s_t$ to capture the  content $\hat{s}_t$ and enumerate feasible actions $\mathcal{A}(o_t)$. State Evaluator \circlednum[response]{2} implements the evaluation function $\mathcal{G}$ by employing the LLM to evaluate the current state of the action sequence using the action sequence $\tau_t$, sequence action memory $M_t$, and experience memory $E$. 
Within a single prompt, the agent (i) summarizes previous interactions, (ii) describes the current state, (iii) determines whether to stop execution early due to significant divergence from the expected state, (iv) decides whether the task is complete, and (v) evaluates whether the intended goal has been successfully achieved.

At each iteration $t$, Action Grounding \circlednum[response]{3} implements action grounding $\pi(.)$ to determine the most suitable next action $a_t$. In our implementation presented in  Algorithm \ref{alg:algorithm1}, the agent first uses the LLM with textual presentation of the current page $\hat{s}_t$, and if there exists two or more applicable actions, the agent uses the image or visual observation $o_t^{\text{img}}$. %This approach reduces the amount of token  

The Rule Extractor \circlednum[response]{4} component implements the long-term experience memory $E$ by extracting rules from previous action sequences.
%\textcolor{red}{~\cite{shinn2024reflexion} show that reflection-based experience can help an LLM improve its performance through reinforcement learning with examples. Specifically, we enhance the central LLM's reasoning performance 
%through a two-phased reflection strategy. This strategy involves 
%by providing the LLM with few-shot examples of correct action sequences and textual rules that summarize past correct or incorrect sequences. These rules are generated by Rule Extractor \circlednum[response]{4} from previous action sequences. For example, a rule might be: ``Ensure to verify the text of the element before clicking, to confirm if it is the desired link and avoid unnecessary clicks on unrelated elements.''}

%Together, 
%This combination allows the LLM to dynamically adjust its approach, improving its ability to predict the most effective next steps and increasing the likelihood of successfully completing the task.

\begin{algorithm}[t]
\footnotesize	
\caption{Iterative LLM-based Agent Planning}
\label{alg:algorithm1}
%\textbf{Input}: a task to complete and an url directing to the intended webpage
\begin{algorithmic}[1] %[1] enables line numbers
%\footnotesize	
%\textbf{Input}: a task to complete and an url directing to the intended webpage
%\begin{algorithmic}[1]
\State Input: $task$ to complete, $url$ of the target webpage, and Experience of previous action sequences $E$ 
%\State $E \gets Experience()$
\State $M_t \gets {\varnothing}$
\Procedure{IterativeAgentPlanning}{$task, url$}
%\State $M_t \gets SequenceActionMemory(task)$
\State $s_t \gets TestEnvironment.load(url)$
% \State $\hat{s}_t, \mathcal{A}(o_t) \gets ContentExtractor(s_t)$ 
\While {$t < MAX_t$}
    \State $\hat{s}_t, A(o_t) \gets ContentExtractor(s_t)$ 

    \State $success, stop \gets \mathcal{G_{LLM}}(\tau_t,M_t, E)$
    
    \If{$success = true$}
        \If{training}
            \State $E \gets E \cup ({M_t, 1})$
        \EndIf
        \State{\textbf{break}}
    \EndIf

    \If{$stop = true$} //early termination
        \If{training}
            \State $E \gets E \cup ({M_t, 0})$
        \EndIf
        \State{\textbf{break}}
    \EndIf
    
    \State $a_t \gets \mathcal{\pi_{LLM}}(\hat{s}_t, \mathcal{A}(o_t), M_t, E)$
    
    \If{count($a_t$) > 1} // $a_t$ is duplicated
        \State $a_t \gets \mathcal{\pi_{LLM}}(o_t^{\text{img}}, a_t, M_t, E)$
    \EndIf
    
    \If{$a_t$ requires input}
        \State $inputvalue \gets  LLM(a_t, M_t, E)$  
        \State $a_t \gets a_t \cup inputvalue$
    \EndIf
    
    \State $M_t \gets M_t \cup (\hat{s}_t, a_t)$
    \State $s_t \gets \mathcal{T} (s_t, a_t)$
    \State $t \gets t + 1$
\EndWhile

\If{$t = MAX_t$}
    \If{training}
        \State $E \gets E \cup ({M_t, 0})$
    \EndIf
\EndIf
    
%\State $E \gets E \cup LLM(\tau_t)$
\State $E \gets E \cup \mathcal{\phi_{LLM}}(\tau_t, E)$ // update rules
\EndProcedure
\end{algorithmic}
\end{algorithm}
\vspace{-6pt} 
\subsection{Test case generation} {\tool} produces test scripts by converting an action sequence that successfully completes a task into a test case, where the task description serves as the test case name. Each step in the test case corresponds to an action executed on a given state. An action is defined by its type (e.g., click, text entry), value (e.g., input text), target web element, and the state of a webpage. The test cases can be automatically replayed in the target environment during regression testing.

%% file: sections/experiment.tex
\section{Experimental Methodology}
\label{sec:eval}

We seek to evaluate (1) the effectiveness of {\tool} in generating action sequences (Sequence Action Effectiveness), (2) its effectiveness in generating test cases (Test Generation Effectiveness), (3) the impact of long-term experience memory, (4) the cost of LLM usage, and (5) the effect of each {\tool} component (Ablation Study).

%We seek to evaluate (1) how effective  {\tool} generates action sequences (Sequence Action Effectiveness), (2) how well  {\tool} in generating test cases (Test Generation Effectiveness), (3) the effect of long-term experience memory, (4)  cost of LLM usage, and effect of each \tool's component. 

%{\bf RQ1. [Sequence Action Effectiveness].} How effective does {\tool} generate action sequences that successfully complete given tasks?

%\noindent {\bf RQ2. [Short-term Memory Analysis].} How accurate does {\tool} perform on the tasks involving multiple dependent web pages?

%{\bf RQ2. [Test Script Generation Effectiveness].} How well is  {\tool} in generating test cases?

%{\bf RQ3. [Experience Analysis].} How well does Experience enhance {\tool}’s performance?

%We will analyze the moving average for Experience, i.e., the average of the correct sequence of actions generated over a specified number of recent runs during the training phase.

%{\bf RQ4. [Ablation Study].} What is the impact of each {\tool}'s component on its performance?

%\noindent {\bf RQ5. [Cost].} What is the cost of using the LLMs?

{\bf Datasets.} We use three datasets. 
First, Real-world dataset (Table~\ref{tab:realworld_data_stat_rq1}), includes 350 tasks from seven popular websites (50 each) generated from task templates (e.g., “Subscribe to the {} channel”).
Second, MiniWoB++~\cite{liu2018reinforcement}, provides 44 sampled tasks with 25 instances each (1,100 total).
Third, OnlineMind2Web~\cite{xue2025onlineMind2Web}, extends Mind2Web~\cite{deng2024mind2web} with 300 tasks from 136 websites across multiple domains. After removing 15 inaccessible ones, 285 tasks from 132 sites were used.

{\color{custom-blue}{
{\bf Baselines.} We restrict our comparison to agents that can emit \textit{executable}, \textit{replayable} action sequences, since this is a prerequisite for test-script generation rather than one-time task execution. Therefore, we chose the following three primary baselines: Li {\em et al.}, SeeAct, and WALT \cite{prabhu2025waltwebagentslearn}. For SeeAct, we adopt $SeeAct_{choice}$, which grounds actions through textual choices. The details on the rationales for selecting these benchmarks and baselines are in the Related Work section.

In addition, Table~\ref{tab:baseline_rq1} includes several other baselines, namely WebNT5, HTML-T5-XL, WebGUM, WGE, CC-Net, RCI, AdaPlanner, and Synapse. These methods are evaluated only on MiniWoB++, as they depend on human demonstrations that are unavailable for the other two datasets.
}}

{\bf Procedure.} We set up Selenium test environment via ChromeDriver. The task in the text string is used as input to {\tool} and baselines. 
%When the task is accomplished, the test scripts and reports are analyzed.
We use \textit{GPT--4o} in our experiments, as it serves as the primary LLM in most baselines and is widely regarded as a state-of-the-art model. 

For MiniWoB++, we first run the training phase for 20 task instances, to obtain the maximum of 8 few-shot examples and a set of textual rules to be used as experience. As in Li {\em et al.}, we evaluated each task on 50 instances. For the real-world dataset, we evaluate on 50 instances for each task.

% The results reported in the original Li {\em et al.} paper are based on \textit{PaLM}, which has been discontinued. Thus, we ran the Le {\em et al.}'s experiment using the codebase with OpenAI's \textit{GPT-4o}.

{\bf Evaluation Metrics}. We define an action as correct if both the selected element (its content) and the predicted operation are correct. For sequence action generation, we use:

{\em Exact Match Rate (\code{Exact-Match})}: We consider the generated SoA for a task as correct only if all actions are correct. \code{Exact-Match} is defined as the ratio of correct results to the total number of results. 

%This metric provides a measure of how effectively the agent completes a given task in multiple scenarios.

{\em Prefix Match Rate (\code{Prefix-Match})}:
\code{Prefix-Accuracy} is the accuracy of the prefix of a generated SoA for a task instance, which is the ratio of continuous correct actions from the start to the first error, divided by the total number of actions. \code{Prefix-Match} is the average in all instances.

%% file: sections/rq1.tex
% \subsection{Effectiveness in Sequence Action Generation (RQ1)}
\subsection{Sequence Action Effectiveness}
\input{sections/rq1-real}
\input{sections/rq1-onlinemind2web}

\input{sections/rq1-mini}
\input{sections/rq1-complexity}

%% file: sections/rq1-real.tex
 \textbf{Performance on \underline{the Real-world dataset}}:
% \label{sec:realworld_exp}
{\color{custom-blue} {
As seen in Table~\ref{tab:realworld_performance_rq1}, 
\tool achieves the highest performance on the Real-world dataset, with 83.8\% Exact-Match and 91.8\% Prefix-Match, outperforming Li {\em et al.} by 56.8\%, WALT by 13.4\%, and SeeAct by 11.4\%. Owing to its iterative planning mechanism (ReAct) and integration of short-term memory, \tool effectively adapts to interface variations and dynamic webpage states. In contrast, Li {\em et al.} employ a static, screen-based planning strategy with full HTML representations, which frequently exceed context limitations and fail to generalize when interfaces change. SeeAct, while demonstrating strong performance on structured platforms such as YouTube, LinkedIn, and Amazon, exhibits a substantial decline on highly dynamic websites like Expedia. Overall, \tool demonstrates superior adaptability and generalization capabilities, maintaining high accuracy even in complex, multi-screen tasks that require sequential reasoning and action execution.
}}
\begin{table}[t!]
\tabcolsep 1.0pt
\centering
\caption{{\color{custom-blue}{Performance of Sequence Action Generation on \underline{Real-world} dataset}}}
\label{tab:realworld_performance_rq1}
\footnotesize
\vspace{-6pt}
\begin{scriptsize} % Các lựa chọn khác: \footnotesize, \scriptsize, \tiny, \large
\begin{tabular}{|c|cccc|cccc|}
\hline
\multirow{2}{*}{\textbf{Website}} & \multicolumn{4}{c|}{\textbf{Exact-Match (\%)}} & \multicolumn{4}{c|}{\textbf{Prefix-Match (\%)}} \\
\cline{2-9}
& \multicolumn{1}{c|}{\textbf{Ours}} & \multicolumn{1}{c|}{\textbf{Li {\em et al.}}} & \multicolumn{1}{c|}{\textbf{SeeAct}} & \textbf{WALT} & \multicolumn{1}{c|}{\textbf{Ours}} & \multicolumn{1}{c|}{\textbf{Li {\em et al.}}} & \multicolumn{1}{c|}{\textbf{SeeAct}} & \textbf{WALT} \\
\hline
YouTube & \multicolumn{1}{c|}{100.0} & \multicolumn{1}{c|}{20.0} & \multicolumn{1}{c|}{100.0} & 100 & \multicolumn{1}{c|}{100.0} & \multicolumn{1}{c|}{20.0} & \multicolumn{1}{c|}{100.0} & 100 \\
Linkedin & \multicolumn{1}{c|}{92.0} & \multicolumn{1}{c|}{46.0} & \multicolumn{1}{c|}{100.0} & 80 & \multicolumn{1}{c|}{98.4} & \multicolumn{1}{c|}{71.0} & \multicolumn{1}{c|}{100.0} & 90 \\
Facebook & \multicolumn{1}{c|}{93.3} & \multicolumn{1}{c|}{33.3} & \multicolumn{1}{c|}{66.7} & 66.7& \multicolumn{1}{c|}{96.7} & \multicolumn{1}{c|}{39.0} & \multicolumn{1}{c|}{79.2} & 75 \\
Google & \multicolumn{1}{c|}{96.0} & \multicolumn{1}{c|}{0.0} & \multicolumn{1}{c|}{100.0} &  60 & \multicolumn{1}{c|}{98.6} & \multicolumn{1}{c|}{0.0} & \multicolumn{1}{c|}{100.0} & 78 \\
 Amazon & \multicolumn{1}{c|}{93.4} & \multicolumn{1}{c|}{54.0} & \multicolumn{1}{c|}{100.0} & 80 &\multicolumn{1}{c|}{94.4} & \multicolumn{1}{c|}{56.0} & \multicolumn{1}{c|}{100.0} & 96 \\
Stackover. & \multicolumn{1}{c|}{80.0} & \multicolumn{1}{c|}{36.0} & \multicolumn{1}{c|}{40.0} & 80 &\multicolumn{1}{c|}{90.0} &  \multicolumn{1}{c|}{45.4} & \multicolumn{1}{c|}{63.3} & 80 \\
Expedia & \multicolumn{1}{c|}{32.0} & \multicolumn{1}{c|}{0.0} & \multicolumn{1}{c|}{0.0} & 20 & \multicolumn{1}{c|}{64.4} & \multicolumn{1}{c|}{9.4} & \multicolumn{1}{c|}{44.1} & 32.7 \\
\hline
Tot./Avg & \multicolumn{1}{c|}{\textbf{83.8}} & \multicolumn{1}{c|}{27.0 } & 72.4 & 69.5  & \multicolumn{1}{c|}{\textbf{91.8 }} & \multicolumn{1}{c|}{34.4 } & \multicolumn{1}{c|}{83.8} & 78.8 \\

% Total/Avg. & \multicolumn{1}{c|}{\textbf{83.8 ($\pm$ 32.6)}} & \multicolumn{1}{c|}{27.0 ($\pm$ 35.8)} & \textbf{72.4 ($\pm$ 42.2)} & \multicolumn{1}{c|}{\textbf{91.8 ($\pm$ 19.8)}} & \multicolumn{1}{c|}{34.4 ($\pm$ 28.3)} & \textbf{83.8 ($\pm$ 43.9)} \\
\hline
\end{tabular}%
\end{scriptsize}
% \vspace{-12pt}
\begin{tablenotes}
  \small
  \item \textit{Note:} \textbf{Ours} denotes our proposed framework, \tool.
\end{tablenotes}
\end{table}

%% file: sections/rq1-onlinemind2web.tex
\textbf{Performance on \underline{OnlineMind2Web}}:
{\color{custom-blue}{Table~\ref{tab:onlinemind2web_performance_rq1} summarizes results on Online-Mind2Web dataset. {\tool} achieves 50.7\% Exact-Match and 62.4\% Prefix-Match over 285 tasks from 132 websites, consistently outperforming Li {\em et al.} and SeeAct across all difficulty levels. Its accuracy reaches 81.3\% on easy, 40.7\% on medium, and 30.0\% on hard tasks.

Compared to the Real-world dataset, Online-Mind2Web includes more complex UI elements such as dropdowns and calendar pickers, where {\tool}’s iterative planning proves effective. It outperforms Li {\em et al.} by 30.5\%, SeeAct by approximately 16\%, and WALT by 4.6\%, particularly on tasks requiring multi-step reasoning and adaptation to dynamic state changes.

Despite these gains, {\tool} still struggles with large interactive elements such as datetime pickers, enter-to-submit fields, and large interactive components that may exceed LLM context limits.
}}

\begin{table}[t!]
\tabcolsep 1.0pt
\centering
\caption{Performance of Sequence Action Generation on \underline{OnlineMind2Web dataset}}
\label{tab:onlinemind2web_performance_rq1}
\footnotesize
\vspace{-6pt}
\begin{scriptsize}
\begin{tabular}{|c|c|cccc|cccc|}
\hline
\multirow{2}{*}{\textbf{Level}} & \multirow{2}{*}{\textbf{No.}} & \multicolumn{4}{c|}{\textbf{Exact-Match (\%)}} & \multicolumn{4}{c|}{\textbf{Prefix-Match (\%)}} \\
\cline{3-10}
&  & \multicolumn{1}{c|}{Ours} & \multicolumn{1}{c|}{Li {\em et al.}} & \multicolumn{1}{c|}{SeeAct} & WALT & \multicolumn{1}{c|}{Ours} & \multicolumn{1}{c|}{Li {\em et al.}} & \multicolumn{1}{c|}{SeeAct} & WALT \\
\hline
\multicolumn{1}{|l|}{Easy} & 80 & \multicolumn{1}{c|}{81.3} & \multicolumn{1}{c|}{40.0} & \multicolumn{1}{c|}{63.8} & 56.3 & \multicolumn{1}{c|}{90.8} & \multicolumn{1}{c|}{54.0} & \multicolumn{1}{c|}{75.8} & 76.6 \\
\multicolumn{1}{|l|}{Medium} & 135 & \multicolumn{1}{c|}{40.7} & \multicolumn{1}{c|}{14.8} & \multicolumn{1}{c|}{30.4} & 36.3 & \multicolumn{1}{c|}{55.0} & \multicolumn{1}{c|}{26.2} & \multicolumn{1}{c|}{42.4} & 68\\
\multicolumn{1}{|l|}{Hard} & 70 & \multicolumn{1}{c|}{30.0} & \multicolumn{1}{c|}{5.7} & \multicolumn{1}{c|}{10.0} & 20 & \multicolumn{1}{c|}{41.2} & \multicolumn{1}{c|}{13.8} & \multicolumn{1}{c|}{20.2} & 62.6 \\
\hline
\multicolumn{1}{|l|}{Tot./Avg} & 285 & \multicolumn{1}{c|}{\bf 50.7} & \multicolumn{1}{c|}{20.2} & \multicolumn{1}{c|}{34.7} & 37.5 & \multicolumn{1}{c|}{\bf 62.4} & \multicolumn{1}{c|}{31.3} & \multicolumn{1}{c|}{46.1} & 69 \\
\hline
\end{tabular}%
\end{scriptsize}
\vspace{-12pt}
\end{table}
% }}

%% file: sections/rq1-mini.tex
\textbf{Performance on \underline{MiniWoB++}}:
% [Main table here - todo for today]
% Another version of main table
%\begingroup
{\color{custom-blue}{
As seen in Table~\ref{tab:miniwob_rq1}, for 975 task instances from 39 tasks, {\tool} achieves 97.4\% Exact-Match and 99\% Prefix-Match. Tasks with fewer than three screens or actions reach near-perfect accuracy (97–100\%), while those with higher complexity slightly decrease to 93\%.

Notably, {\tool} attains performance comparable to leading baselines without human demonstrations—matching WebGUM (401K demos) and surpassing CC-Net (BC + RL) (2.4M demos). Although Synapse achieves perfect accuracy, it depends on 154 handcrafted exemplars and task-specific prompts, which risk overfitting by reducing the need for multi-state exploration.
}}
\begin{table}[t]
\footnotesize
\tabcolsep 4pt % Tăng nhẹ tabcolsep để giãn khoảng cách cột cho đẹp
\centering
\caption{Action Sequence Generation on \underline{MiniWoB++}. Exact-match (\%), Correct runs (975 total).}
\label{tab:miniwob_rq1}
\vspace{-6pt}
\begin{scriptsize}

\begin{tabular}{|l|cccc|}
\hline
\textbf{Metrics} & \textbf{Li (4o)} & \textbf{SeeAct} & \textbf{Ours} & \textbf{WALT} \\ \hline 
\textbf{Exact-Match (\%)} & 91.7\% & 69.9\% & 97.4\% & 86.9\% \\ \hline
% \textbf{\# Correct runs}   & (949)    & 682    & 950    & 847    \\ \hline
\end{tabular}
\end{scriptsize}
\end{table}

\begin{table}[t]
\centering
\footnotesize
%\vspace{-9pt}
\caption{Performance on 125 Challenging Task Instances (MiniWoB++).}
\label{tab:additional_miniwob_RQ1}
\vspace{-7pt}
\begin{scriptsize} % Các lựa chọn khác: \footnotesize, \scriptsize, \tiny, \large

\begin{tabular}{|c|cccc|}
\hline
\multirow{2}{*}{\textbf{Tasks}} & \multicolumn{4}{c|}{\textbf{Exact-Match (\%)}} \\ \cline{2-5}
 & \multicolumn{1}{c|}{\textbf{Ours}} & \multicolumn{1}{c|}{\textbf{Li {\em et al.}}} & \multicolumn{1}{c|}{\textbf{SeeAct}} & \multicolumn{1}{c|}{\textbf{ WALT }} \\ \hline
choose-date & \multicolumn{1}{c|}{92.0} & \multicolumn{1}{c|}{24.0} & \multicolumn{1}{c|} {0.0}  & \multicolumn{1}{c|}{0.0}\\
book-flight & \multicolumn{1}{c|}{100.0} & \multicolumn{1}{c|}{24.0} & \multicolumn{1}{c|}{0.0}&\multicolumn{1}{c|}{0.0}\\
flight.AA   & \multicolumn{1}{c|}{100.0} & \multicolumn{1}{c|}{0.0}  & \multicolumn{1}{c|}{40.0} &\multicolumn{1}{c|}{0.0}\\
flight.Alaska & \multicolumn{1}{c|}{96.0} & \multicolumn{1}{c|}{0.0}  & \multicolumn{1}{c|}{0.0} & \multicolumn{1}{c|}{0.0}  \\
flight.Alaska-auto & \multicolumn{1}{c|}{20.0} & \multicolumn{1}{c|}{0.0} & \multicolumn{1}{c|}{0.0}  & \multicolumn{1}{c|}{0.0}   \\ \hline
Total/Avg.  & \multicolumn{1}{c|}{81.6 ($\pm 34.6$)} & \multicolumn{1}{c|}{9.6 ($\pm 13.1$)} &  \multicolumn{1}{c|}{8.0 ($\pm 18$)} &  \multicolumn{1}{c|}{0.0} \\ \hline
\end{tabular}
\end{scriptsize}
\end{table}

\color{custom-blue}{{
Among methods that do not rely on human demonstrations, {\tool} slightly outperforms Li {\em et al.} in \code{Exact-Match} accuracy (97\% vs. 94\%), while surpassing SeeAct by 20\% and WALT by 10.5\%. On 125 challenging MiniWoB++ tasks (Table~\ref{tab:additional_miniwob_RQ1}) involving an average of five screens and seven actions, {\tool} achieves 81.6\% Exact-Match, compared to 10\% for Li {\em et al.} and 8\% for SeeAct. WALT struggles to perform these tasks. 

As shown in Table~\ref{tab:miniwob_rq1}, SeeAct attains 69.9\% Exact-Match overall but falls to 8\% on complex tasks, limited to the \code{flight.AA} case. Its failures arise from difficulty handling datetime pickers, dropdowns, and icon-only buttons. In contrast, {\tool} exhibits superior robustness and accuracy on multi-step, dynamic tasks.
}}

%% file: sections/rq1-complexity.tex
\begin{figure}[t!]
    \centering
    % Hình bên trái
    \begin{subfigure}[b]{0.48\textwidth}
        \centering
    \includegraphics[width=\textwidth]{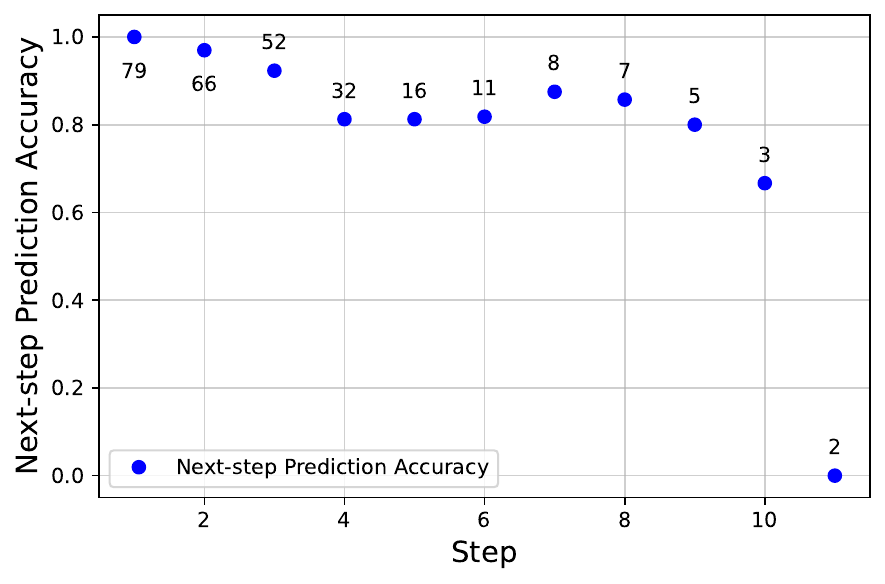}
        \caption{Next-step Prediction.}
        \label{fig:step-acc}
    \end{subfigure}
    \hfill % Tạo khoảng trống tối đa ở giữa để đẩy 2 hình ra 2 bên
    % Hình bên phải
    \begin{subfigure}[b]{0.48\textwidth}
        \centering
        \includegraphics[width=\textwidth]{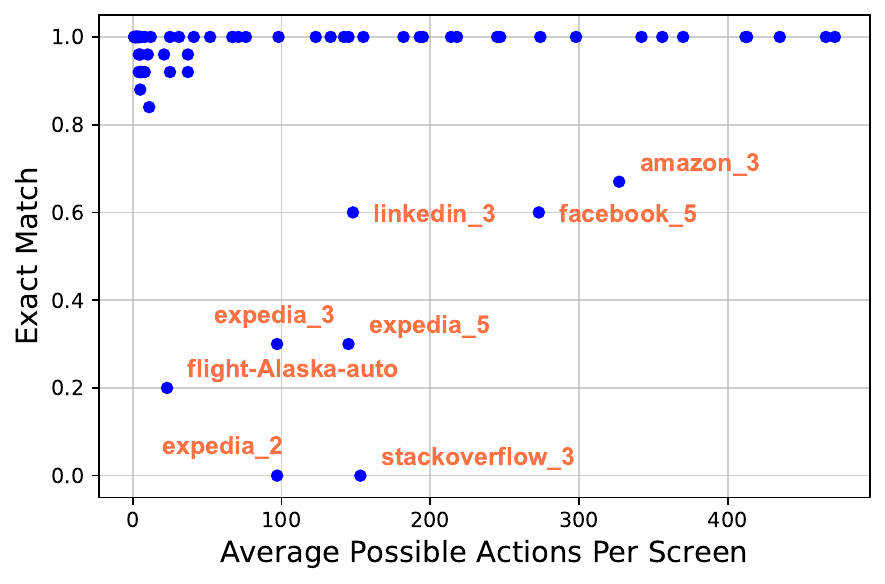}
        \caption{Accuracy by Actions/Screens.}
        \label{fig:action-screen-acc}
    \end{subfigure}
\vspace{-21pt}
    \caption{Performance by Task Complexity}
    \label{fig:experience_rq2}
\end{figure}

\textbf{Accuracy by Task Complexity}:
{\color{custom-blue}{
We evaluate {\tool} by task complexity, measured through action sequence length and actions per screen.

{\em Next-step prediction accuracy by SoA's length}.
As seen in Fig.~\ref{fig:step-acc}, {\tool} achieves 100\% first-step accuracy on MiniWoB++ and Real-world datasets, with a gradual decline as sequence length increases. A similar trend~occurs in Online-Mind2Web, where \code{Exact-Match} drops from 81.3\% (easy) to 40.7\% (medium) and 30.0\% (hard).

{{\em Exact-Match sequence accuracy by number of actions per screen}}. 
In Fig.~\ref{fig:action-screen-acc}, {\tool} maintains high \code{Exact-Match} even with up to 500 actions per screen. Performance decreases on long or dynamic tasks (e.g., pop-ups, date pickers, complex forms), with some outliers below 70\%, mainly due to hidden options or invalid data generation.
}}

%% file: sections/rq-test.tex
\subsection{Test Generation Effectiveness}

\begin{table}[t]
\centering
\footnotesize
\caption{{\color{my-blue}{Success Rates of Generated Test Scripts}}}
\label{tab:test-gen-rq5}
\vspace{-6pt}
\begin{scriptsize}

\begin{tabular}{|c|l|}
\hline
\textbf{Test scripts for} & \textbf{Valid / Successful Trial} \\ \hline
MiniWoB++                &     1048/1048 = 100.0\%                               \\
Real-world Application    &     276/304 = 90.8\%                               \\
OnlineMind2Web &     {\color{my-blue}{109/141 = 77.3\%}}                               \\ \hline
\end{tabular}%
\end{scriptsize}
\vspace{-12pt}
\end{table}

{\color{custom-blue}{
We used {\tool} to generate regression test scripts for the {\color{my-blue}{1,493}} tasks with correct SoAs and executed them on the target web applications. A script is considered valid if it (1) executes all steps without errors and (2) successfully reaches the intended goal. As shown in Table~\ref{tab:test-gen-rq5}, all successful SoAs in MiniWoB++ produced valid scripts, while 90.8\% of successful SoAs from real-world applications executed correctly. The higher reliability in MiniWoB++ stems from its static, controlled environment, whereas real-world sites introduce dynamic content and runtime variability—for example, layout changes or fluctuating search results may invalidate XPath locators. On LinkedIn, the “Jobs” button may yield different XPaths across layouts, and on Expedia, network delays occasionally caused actions to execute before elements fully loaded, resulting in failures.
}}
% We used {\tool}  to generate test scripts for regression testing for the {\color{my-blue} {1,493}} tasks with correct sequences of actions in RQ1. We then executed them on the Web applications under test. Although test scripts are interpreted one-to-one from the SoA, they may fail during execution due to various reasons. Thus, a test script is valid if it (1) executes all~steps~without errors, and (2) successfully achieves the intended testing goal. 
% As seen in Table~\ref{tab:test-gen-rq5}, all successful SoAs in the MiniWoB++ dataset produced valid test scripts. For the real-world applications, 90.8\% of successful SoA led to valid test scripts. Simulated MiniWoB++ environment comprises static, controlled HTML pages, making execution more predictable. In contrast, real-world applications often involve dynamic content and runtime variability, which can cause scripts to fail. For example, changing layouts or search result content, which can lead to inconsistent/invalid XPath locators. For LinkedIn, the same "Jobs" button may produce different XPath locators depending on the page's layout during action sequence generation, resulting in invalid tests. For Expedia, where fluctuating network latency caused the test script to attempt a click flight before the search results were fully loaded, leading to failure. 

%Improving test script robustness is beyond our scope. However, approaches exist for addressing this problem, such as generating robust locators~\cite{nguyen2021generating,leotta2016robula+}, or selecting relevant locators~\cite{nass2023similarity}. 

{\color{custom-blue}{On the OnlineMind2Web dataset, our approach achieves a 77.3\% success rate. Although many tasks resemble those in the Real-world dataset, the broader variety of website types introduces failures on pages with dynamic banners, dropdowns, or frequently changing navigation paths. For example, when searching “SAM2” on GitHub, the generated script may succeed during execution but later return a \textit{429 Too Many Requests} error due to dynamic request limits.

Techniques such as generating more robust locators~\cite{nguyen2021generating,leotta2016robula+} or selecting relevant ones~\cite{nass2023similarity} may mitigate these issues.
}}
% {\color{my-blue}{On the OnlineMind2Web dataset, a success rate of 77.3\% is achieved. While this dataset has many tasks similar to those of the Real-world dataset, it also covers a diverse range of website categories, and failures often occur on sites with dynamic banners, dropdowns with dynamic content, or frequently changing navigation paths. For example, when searching for “SAM2” on GitHub, the generated script successfully reaches the listing page during execution. However, during evaluation, the same script may instead return a \textit{429 Too Many Requests} HTTP request error, an instability caused by dynamic request limits.}} 
% Solutions such as generating robust locators~\cite{nguyen2021generating,leotta2016robula+} or selecting relevant ones~\cite{nass2023similarity} may help mitigate these issues.

%Improving the robustness of generated test scripts is beyond the scope of this work. Nevertheless, 

%Future research should focus on improving test script robustness, e.g., utilizing multiple interchangeable locators (e.g., CSS selectors, element attributes, text), or incorporating waiting strategies or dynamic synchronization to better handle runtime variability.

%% file: sections/rq3.tex
% \vspace{9pt}
\subsection{Experience Analysis}

% 3 main points need to be made:

% 1. Model learns more correct knowledge during our training. 

% 2. Using rules better than not using

% 3. After getting to 1, the interaction with humans can be stopped because experience becomes "perfect".

% \begin{figure}[t]
% \centering
% \includegraphics[width=3in]{figures/RQ3_MovingAvg.png} %example1.png
% \caption{Examples of Task Descriptions and Web Pages}
% \vspace{-2mm}
% \label{fig:motiv1}
% \end{figure}

% Bo moving average, bo runs, cho no bu len, extend cho khoang trang ben trai phai

% Excel thanh 4 cai minh, roi lam minipage tren duoi 4 cai hinh, 2 tren 2 duoi.
%\vspace{-12pt}
\begin{figure}[h]
    \centering
    % --- Row 1 ---
    \begin{subfigure}[t]{0.45\columnwidth}
        \centering
        \includegraphics[width=\linewidth]{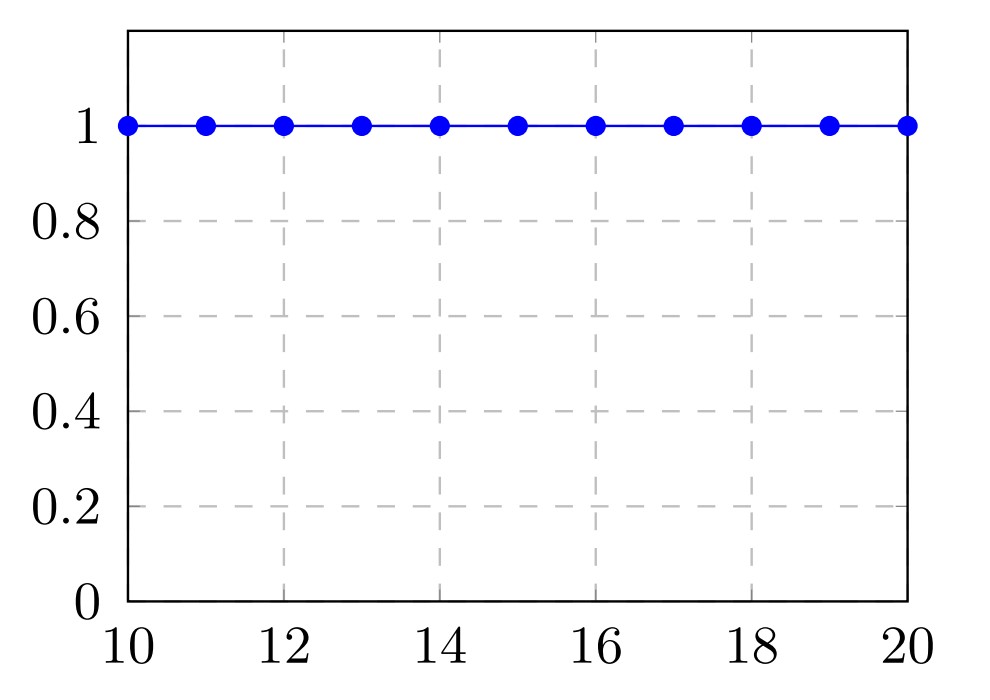}
        \caption{click-button}
        \label{fig:click-button}
    \end{subfigure}
    \hfill
    \begin{subfigure}[t]{0.45\columnwidth}
        \centering
        \includegraphics[width=\linewidth]{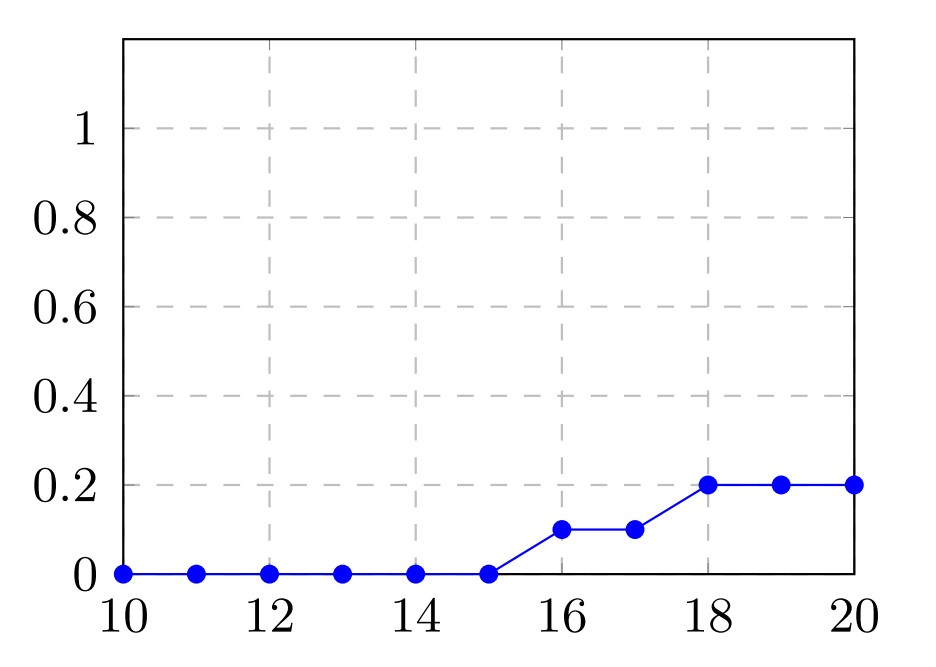}
        \caption{Alaska-auto}
        \label{fig:alaska-auto}
    \end{subfigure}

    \vspace{6pt} % Space between rows

    % --- Row 2 ---
    \begin{subfigure}[t]{0.45\columnwidth}
        \centering
        \includegraphics[width=\linewidth]{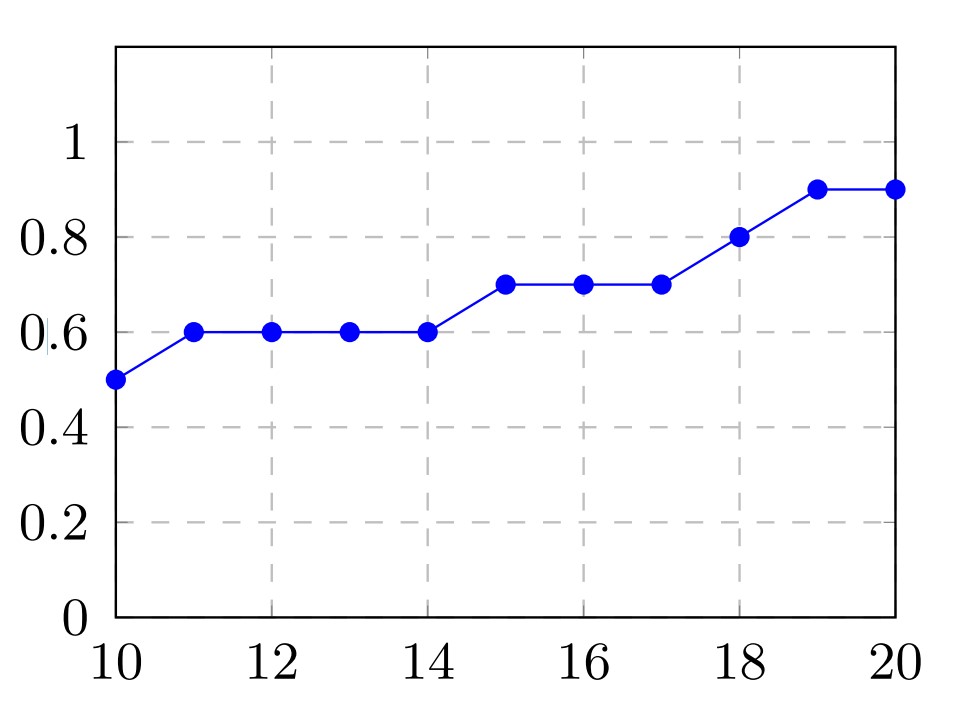}
        \caption{search-engine}
        \label{fig:search-engine}
    \end{subfigure}
    \hfill
    \begin{subfigure}[t]{0.45\columnwidth}
        \centering
        \includegraphics[width=\linewidth]{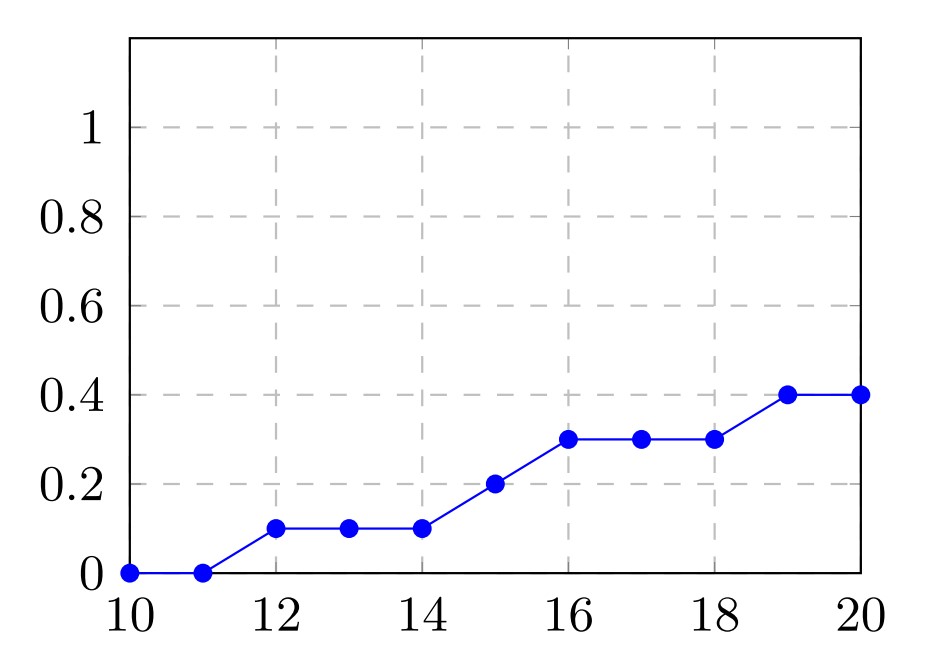}
        \caption{search w/o rules}
        \label{fig:search-engine-w/o-rule}
    \end{subfigure}

    \vspace{-6pt}
    \caption{Moving average of selected tasks in training}
    \vspace{-12pt}
    \label{fig:experience_rq2}
\end{figure}

% \begin{figure}[h]
%     \centering
%     \begin{subfigure}[t]{0.24\columnwidth}
%         \centering
%         \includegraphics[width=\columnwidth]{figures/click_button_moving_avg_rq2.jpg}
%         \caption{click-button}
%         \label{fig:click-button}
%     \end{subfigure}%
%     \begin{subfigure}[t]{0.24\columnwidth}
%         \centering
%         \includegraphics[width=\columnwidth]{figures/Alaska_auto_moving_avg_rq2.jpg}
%         \caption{Alaska-auto}
%         \label{fig:alaska-auto}
%     \end{subfigure}
%     \begin{subfigure}[t]{0.24\columnwidth}
%         \centering
%         \includegraphics[width=\columnwidth]{figures/search_engine_moving_avg_rq2.jpg}
%         \caption{search-engine}
%         \label{fig:search-engine}
%     \end{subfigure}
%     % \vspace{-6pt}
%     \begin{subfigure}[t]{0.24\columnwidth}
%         \centering
%         \includegraphics[width=\columnwidth]{figures/search_engine_no_rule_moving_avg_rq2.jpg}
%         \caption{search w/o rules}
%         \label{fig:search-engine-w/o-rule}
%     \end{subfigure}
%      \vspace{-6pt}
%     \caption{Moving average of selected tasks in training (RQ3).}
%     \label{fig:experience_rq2}
% \end{figure}

Fig.~\ref{fig:experience_rq2} shows the moving average for Experience, i.e., the average correct SoA over a window of 10 recent runs during training. Comparing Figures~\ref{fig:search-engine} and \ref{fig:search-engine-w/o-rule} (without experience rules), the moving average starts at 0.5 (5 correct sequences over 10 runs) with rules but at 0 without. Consequently, after 20 runs, the agent {\em without experience rules} reaches only 0.4, versus 0.9 ({\em 9 correct sequences over the last 10 runs}) when rules are included. This rising trend indicates that {\tool} keeps acquiring correct knowledge during training. For \code{click-button}, with the best possible initial performance (Fig.~\ref{fig:click-button}), retention stays consistent at a 100\% moving average, whereas for a complex task (\code{Alaska-auto}) it is lower, as the correct sequence is harder to obtain. Tracking the moving average monitors the quality of knowledge in experience memory and helps select the best few-shot examples and textual rules for prediction: we can halt training once it exceeds a threshold (Fig.~\ref{fig:search-engine}) and capture the knowledge at that point in the Experience prompts for subsequent predictions.

% Figure 6 shows the moving average
% calculated by the proportion of successful episodes over a window size of 10 episodes. There are three types of scenarios typically encountered when apply-
% ing the training phase. The first type was for the easier tasks (click-button) that the agent can do well without the training phase. These tasks could be easily
% identified through their high initial accuracy in the training phase of around 100\% (6a). Some tasks, however, gained many benefits from the accumulated knowledge acquired during the training phase; search-engine, for example, improved from 50\% to 90\% in success rate (6b). Finally, there were also challenging tasks, like Alaska-auto (6c) whose success rate remained at 20\% at the end of the
% training, that failed to achieve decent accuracy even with the training phase.

%% file: sections/rq4.tex
\subsection{Ablation Study}

%As seen in Table \ref{tab:rq4},

%including Experience, Sequence Actions Memory, and Iterative Planning in {\tool}, 

{\em Ablation Study on Key Components.} We removed {\tool}'s key components and compared their performance. For each component, we removed the corresponding section in the main prompt.
%We removed the Experience component from \tool by specifically removing the experience portion (Figure~\ref{prompt:experience_prompt}) from the main prompt (Figure~\ref{prompt:iterative_prompt}) to create the variant \textit{w/o experience}. 
%For the variant \textit{w/o sequence actions memory (SAM) and experience}, we removed both SAM (Section~\ref{sec:seq_action_mem}) and Experience by removing the sections for those two components from the main prompt. 
%
However, for the variant \textit{w/o iterative planning}, we changed the main prompt to ask LLM only once to generate a complete SoA without the loop for planning. As seen in Table \ref{tab:rq4}, all variants have lower accuracies than {\tool}, indicating the contributions of each component. 
Without sequence actions memory, the variant \code{w/o (SAM} + \code{experience)} has the most drop in \code{Exact-Match} (47 vs. 97).
Thus, SAM is most important with its retaining the history of previous steps and evolving states to decide the next~action. We found that \code{w/o (SAM} + \code{experience)} kept repeating some action multiple times until the step limit.  {\tool} \textit{w/o iterative planning} has a decrease of 23\% in \code{Exact-Match} where~it struggles on tasks with more screens. The variant \textit{w/o experience} has the smallest drop of 9\% in \code{Exact-Match}. This confirms our idea that experience with reflection in Li {\em et al.} has less impact than the combination of  \code{SAM+experience} and iterative planning.

\begin{table}[t]
\centering
\footnotesize
\caption{Contribution of Different Components}
\vspace{-6pt}
\label{tab:rq4}
%\resizebox{\columnwidth}{!}{%
\begin{scriptsize}

\begin{tabular}{|l|c|c|}
\hline
\textbf{Module}                                   & \multicolumn{1}{|c|}{\textbf{\code{EM} (\%)}} & \multicolumn{1}{|c|}{\textbf{\code{PM} (\%)}} \\ \hline
\textbf{\tool} (4o)                                  & \textbf{97}        & \textbf{99}        \\ \hline
\textit{w/o experience}                           & 88                 & 93                 \\
\textit{w/o (SAM + experience)} & 47                 & 67                 \\
\textit{w/o iterative planning}             & 75                    & 85                    \\ \hline
\end{tabular}%
%}
\end{scriptsize}
\vspace{-12pt}
\end{table}

%\endgroup

%\begingroup
%\setlength{\tabcolsep}{5pt} % Default value: 6pt
%\renewcommand{\arraystretch}{1.5} % Default value: 1
%\begin{table}[t]
%\centering
%\caption{Ablation study on memory capacities.}
%\label{tab:ablate_memory}
%\begin{tabular}{|l|c|c|}
%\hline
%\textbf{Module}                                   & \multicolumn{1}{|c|}{\textbf{\code{Exact-Match} (\%)}} & \multicolumn{1}{|c|}{\textbf{\code{Prefix-Match} (\%)}} \\ \hline
%\textbf{Memory size All}                                  & \textbf{97}        & \textbf{99}        \\ %\hline
%\textit{Memory size 3}                           & 94                 &       96           \\
%\textit{Memory size 1} & 91                 &       94           \\
%\hline
%\end{tabular}%
%\end{table}
%\endgroup

{\em Ablation Study on Varying Sizes of Short-term Memory}. Our results show that restricting memory to {\em only 1--3 most recent actions} reduces \code{Exact-Match} by 6\% and 3\%, respectively. Full access to short-term memory proves beneficial for tasks where the current state does not fully reflect the completion of prior steps.
%or for complex form-filling tasks involving multiple fields. 
For example, in the \code{click-tab-2} task, full memory access enables \tool to track all visited tabs, preventing redundant clicks on the same tab, an issue with limited memory.

%% file: sections/rq6.tex
\subsection{Cost Analysis}
\label{sec:cost}
% Analyze token consumption
% Evaluate token for common case in real-wolrd application
% Analyze Token for each component
% The repetition of token
% Ways to optimize or improve

% Analyze case that consume significant token, comment the linear/ growth

Figure~\ref{fig:token_main} shows the token distributions in prompts and outputs for each {\tool} component: \textit{content tokens} (Content Extractor), \textit{agent tokens} (Action Grounding), and \textit{rule tokens} (Rule Extractor). These components are invoked repeatedly during the iterative loop, and prompt token counts vary with the complexity of the web application and the action-sequence length. Action Grounding consumes the most tokens, followed by Content Extractor, as both manage the web state and are critical to the iterative process. Figure~\ref{fig:token_input} presents token consumption for input generation in Action Grounding, executed once per input field in the task. Overall, the average cost across all tasks is 80k input and 22k output tokens for both training and evaluation.

Compared with WALT, {\tool} consumes about one-third of WALT's total tokens; for output tokens, the gap is larger still, at only 8.9\% of WALT's count (see Section \ref{sec:CostAnalsysis}).

%\begin{figure}[t!]
%    \centering
%    \includegraphics[width=2.5in]{figures/token_usage.pdf} 
%    \vspace{-6pt}
%    \caption{\#tokens for each component's input/output (log scale).}
%    \label{fig:token_main}
%\end{figure}

%\begin{figure}[t!]
%    \centering
%    \includegraphics[width=3in]{figures/token_usage_input.pdf} 
%    \vspace{-6pt}
%    \caption{\#tokens for input generator (log scale).}
%    \label{fig:token_input}
%\end{figure}

% tăng font chữ lớn hơn và bỏ chữ tokens

% Scale trong ngu canh ung dung phuc tap, phan tich chi phi, scalabiliy in the loop

% \begin{figure}[htpb]
%     \centering
%     \includegraphics[width=0.6\textwidth]{figures/token_sequence.png} 
%     \caption{Number of tokens for each components input/output (log scale).}
%     \vspace{-2mm}
%     \label{fig:overview}
% \end{figure}

%% file: sections/related.tex
\paragraph{{\bf Benchmarks for Web Task Completion and UI Agents.}} 
The rapid progress of web UI agents has been supported by realistic benchmarks for navigation and interaction, such as WebArena~\cite{zhou2023webarena}, 
WebLINX~\cite{lu2024weblinx}, 
WebGames~\cite{webgames2025}, 
OSUniverse~\cite{osuniverse2025}, BEARCUBS~\cite{bearcubs2025}, and EconWebArena~\cite{econwebarena2025}
BrowseComp~\cite{browsecomp2025}.
Since these benchmarks are defined in natural language rather than executable form, we collected our own real-world dataset (Table~\ref{tab:realworld_data_stat_rq1}) and adopted MiniWoB++~\cite{liu2018reinforcement} and OnlineMind2Web~\cite{xue2025illusionprogressassessingcurrent}, which better align with test generation using available or {\bf executable} sequences of actions.

\textbf{Web Task Completion and Web UI Agents.} Several approaches have been proposed to
build task completion and UI agents including Browsing~\cite{browsing2025}, ShowUI~\cite{showui2025}, OpenAI’s Operator~\cite{operator2025}, SeeClick~\cite{cheng-etal-2024-seeclick}, CogAgent~\cite{hong2024cogagentvisuallanguagemodel}, InfiGUIAgent~\cite{infiguiagent2025}, UI-TARS~\cite{qin2025uitarspioneeringautomatedgui}, UI-AGILE~\cite{ui_agile2025}, UI-Venus~\cite{ui_venus2025}.
%Recent systems integrate diverse strategies: Beyond Browsing~\cite{browsing2025} mixes browser actions with API calls, ShowUI~\cite{showui2025} grounds screenshots for zero-shot generalization, and OpenAI’s Operator~\cite{operator2025} demonstrates industrial-scale multimodal browser control. {\em Yet these agents only report task completion and do not explicitly produce SoAs, therefore, we did not compare them with {\tool} in test generation}.

%\textbf{Web UI Agents.} Multimodal UI agents combine DOM structures with screenshots to improve grounding (e.g., SeeClick~\cite{cheng-etal-2024-seeclick}, CogAgent~\cite{hong2024cogagentvisuallanguagemodel}). Recent advances include reflection-based reasoning (InfiGUIAgent~\cite{infiguiagent2025}), iterative screenshot-only interaction (UI-TARS~\cite{qin2025uitarspioneeringautomatedgui}), reinforcement learning strategies (UI-AGILE~\cite{ui_agile2025}), and trajectory alignment (UI-Venus~\cite{ui_venus2025}). 

\textbf{Computer-Using Agents.} Beyond the web, agents have been developed for mobile (Guardian~\cite{ran2024guardian}, GPTDroid~\cite{gptdroid}, AutoDroid~\cite{Wen2023AutoDroidLT}, DroidAgent~\cite{yoon2023autonomouslargelanguagemodel}) and desktop use (TaskMatrix.AI~\cite{liang2024taskmatrix}, Computer Use~\cite{gur2023computer}, Claude 3.5 Sonnet~\cite{claude_computer_use2025}). 

However, most of these agents are designed to perform computer-using tasks with LLMs that represent actions as absolute screen coordinates or natural-language descriptions. Such representations are brittle for regression testing and, as designed, may not be executed properly without invoking an LLM at replay time. In contrast, our approach generates scripts that run regression tests without any subsequent LLM involvement. We therefore exclude these agents from our comparison: their outputs (textual descriptions or absolute coordinates) would first need to be converted into executable actions for test generation, a process that could introduce errors and lead to unfair comparisons. Instead, we adopt SeeAct~\cite{pmlr-v235-zheng24e} and WALT~\cite{prabhu2025waltwebagentslearn} as our two baselines, since {\em both can prompt LLMs to output executable SoAs} and are recent state-of-the-art approaches.

%These systems extend LLM control to apps, files, and OS-level interactions, but like web task completion agents, {\em they focus on accomplishing tasks rather than producing machine-executable SoAs for automated testing} as in {\tool}.

%Cucumber \cite{cucumber} with Gherkin was an early approach using texts for test flow creation,  requiring manual step definitions. Later methods automatically generated test paths from navigation models~\cite{biagiola2019diversity,zheng2021automatic,sherin2023qexplore} or human execution traces~\cite{ermuth2016monkey,paiva2020test}. KeyjaxTest~\cite{qi2019leveraging} introduced keyword-guided exploration for targeted path generation. 

% \textbf{LLM-based Mobile Testing.} LLMs enabled applications in mobile testing~\cite{liu2024make,wen2024autodroid,liu2023fill,liu2024testing,ran2024guardian}, unit testing~\cite{lemieux2023codamosa,rao2023cat,dakhel2024effective,yuan2024evaluating}, test oracles~\cite{nashid2023retrieval}, web agents~\cite{thil2024navigating,hong2024cogagent,cheng2024seeclick,lai2024autowebglm,chen2024webvln,he-etal-2024-webvoyager}. 

\textbf{Relation to reflection-based agents.} \tool shares ingredients with ReAct~\cite{yao2023react} and Reflexion~\cite{shinn2024reflexion} but differs in its control structure. ReAct interleaves reasoning and acting without persistent cross-episode memory and Reflexion repairs reactively, revising only after a full trajectory fails. \tool instead (i) separates \textbf{state evaluation} from \textbf{action grounding} via a dedicated evaluator $\mathcal{G}_{\text{LLM}}$
 that decides completion or early termination after every step, enabling \textbf{proactive} correction rather than end-of-trajectory repair; (ii) extracts structured behavioral rules incrementally from both successful and failed trajectories during training; and (iii) conditions each subsequent action on this accumulated, abstracted rule set rather than on raw trajectory replay alone. The result is a self-improving control loop with persistent experience abstraction, and the ablation in Section 5.4 quantifies the payoff: removing experience drops Exact-Match by 9 points, and removing short-term memory with it drops by 50 points, confirming that the combination improves performance.

%% file: sections/conclusion_future_work.tex
% In automated web testing creating action sequences from natural-language task descriptions for future test scripts is crucial. Current approaches either require substantial manual effort or do not consider the history of previous web content and actions. This paper introduces {\tool}, an iterative LLM-based agent planning approach that determines the next action based on observations of current content, short-term memory of previous states and actions, and long-term experience with (in)correct sequences. Our evaluation shows that {\tool} achieves high performance on complex tasks
% %, with 88\% of action sequences exactly matching the ground truth, 
% and outperforms the baselines on all tasks without much manual effort.

%In automated web testing, creating action sequences from natural-language task descriptions for future test scripts is crucial. Current approaches either require substantial manual effort or do not consider the history of previous web content and actions. 

%This paper presents {\tool}, an iterative LLM-based agent planning approach that selects actions using current content, short-term memory, and long-term experience. {\tool} achieves 97\% exact match on complex tasks and performs better than top baselines on MiniWoB++ tasks without manual effort. On real-world apps, it reaches 87\% exact and 93\% prefix match, outperforming baselines by 59\%. 

We present {\tool}, a multi-modal, iterative Web testing agent that uses vision-capable LLMs to generate test scripts for GUI tasks described in natural language. We showed that long-term experience memory, with rules extracted from past experience, helps the agent make better decisions, and that grounding the model's reasoning in the actual text and visual context at each step reduces hallucinations. Moreover, {\tool}'s proactive correction reassesses the state after each step to adjust its plan, in contrast to state-of-the-art self-reflection approaches, whose reactive repair attempts corrections only after an entire trajectory has failed. Our evaluation on three datasets shows superior performance over the baselines.

This work suggests promising directions for test generation and autonomous end-to-end testing. As LLMs and multimodal reasoning advance, future frameworks may not only generate reliable action sequences but also autonomously perform tests, reducing human effort. Research may move toward fully autonomous, self-reflective agents that continuously learn to test and verify diverse applications without human intervention—pointing to a future in which AI-driven testing agents act as trustworthy collaborators, improving software reliability while reducing human~effort.

%\section{Data Availability}

\vspace{2pt}
\noindent {\bf Data Availability}. Data/code are available on our website~\cite{project-website}.

%% file: sections/threats-to-validity.tex
\subsection*{Limitations}
{\color{custom-blue}{
% 
% Several factors may affect this study’s validity. A key internal threat is LLM hallucination, as outputs can vary across runs \cite{Huang_2025}. We mitigated this by setting the temperature to zero and using focused prompts, resulting in consistent outputs.

% Regarding external validity, the three datasets may not cover all web app types or UI behaviors. However, MiniWoB++ is a standard benchmark (e.g., \cite{liu2018reinforcement, li2023zero, zheng2023synapse, thil2024navigating, furuta2023multimodal}), Online-Mind2Web spans 136 real-world websites, and our curated dataset includes diverse modern apps such as Facebook, LinkedIn, Amazon, and StackOverflow. Together, they offer a balanced and representative evaluation of both controlled and real-world web environments.

% old
There are several concerns that might affect the validity of this study. One internal threat to validity is concerned with the hallucination of the LLMs as their outputs are not guaranteed to be consistent between different runs \cite{Huang_2025}. To alleviate this concern, we set the temperature parameter, which affects the randomness of generated responses of the LLMs, to zero to produce a more consistent and deterministic output. We defined the prompts to be focused and to request specific outputs. As a result, our experiments showed consistency in the generated outputs by the LLMs. 

The use of three datasets may limit the generalizability of our study. The applications in those datasets might not be representative of Web apps, and their UI elements and interactions may not reflect all types of elements found in general applications. Nonetheless, the MiniWoB++ dataset is widely used as a benchmark for automating computer tasks in previous studies, e.g., \cite{liu2018reinforcement, li2023zero, zheng2023synapse, thil2024navigating, furuta2023multimodal}. The Online-Mind2Web includes tasks from 136 real-world websites, covering diverse scenarios that users perform online. Seven real-world applications in our dataset are popular from various domains from social networks (Facebook, LinkedIn) to e-commerce (Amazon) to professional support (Stackoverflow). They are modern and well-updated applications that use common and diverse sets of UI elements. They include UI elements and actions that are typically found in modern web-based applications today. Taken together, the combination of MiniWoB++, Online-Mind2Web, and our curated dataset provides both controlled dynamic tasks and realistic real-world applications, offering a broad and meaningful representation of modern web environments.
}}

Our evaluation is Web-specific: \tool operates over DOM elements and executes actions in a Selenium test environment, and we do not empirically demonstrate transfer to mobile or desktop platforms. The design, however, separates the interface representation from the control logic. The Content Extractor produces a structured state, a set of feasible actions over interactable elements together with the surrounding UI context, which in the Web setting is derived from the DOM, but the same abstraction is available on other platforms: Android exposes a View Hierarchy through accessibility APIs, and desktop environments expose Accessibility Trees (e.g., via UI Automation or AT-SPI), both of which provide structured nodes with attributes such as type, text, and interactability. Action Grounding would then select an executable element and operation from these structured nodes rather than from DOM elements. Because \tool's higher-level components including state evaluation, rule extraction, and experience-conditioned action grounding, depend only on the availability of a structured UI representation and not on Web-specific features, the framework is platform-agnostic at the architectural level. Empirical validation on Android and desktop is a promising direction for extending end-to-end testing across platforms.

%% file: sections/motiv-icse25.tex
\section{Motivating Example}

%\subsection{Example}
% Introduction of search-engine task

%Let us use examples to illustrate the problem and motivate our solution. 

%Fig.~\ref{fig:motiv2}. The task description states, {\em``Use the textbox to enter `Macie' and press `Search', then find and click on the 8th search result.''}

%\begin{figure}[t]
%\centering
%\includegraphics[width=3in]{figures/example2.png} %example1.png
%\caption{Examples of task descriptions and webpage.}
%\vspace{-2mm}
%\label{fig:motiv1}
%\end{figure}
\begin{figure*}[t]
\centering
\includegraphics[width=0.9\textwidth]{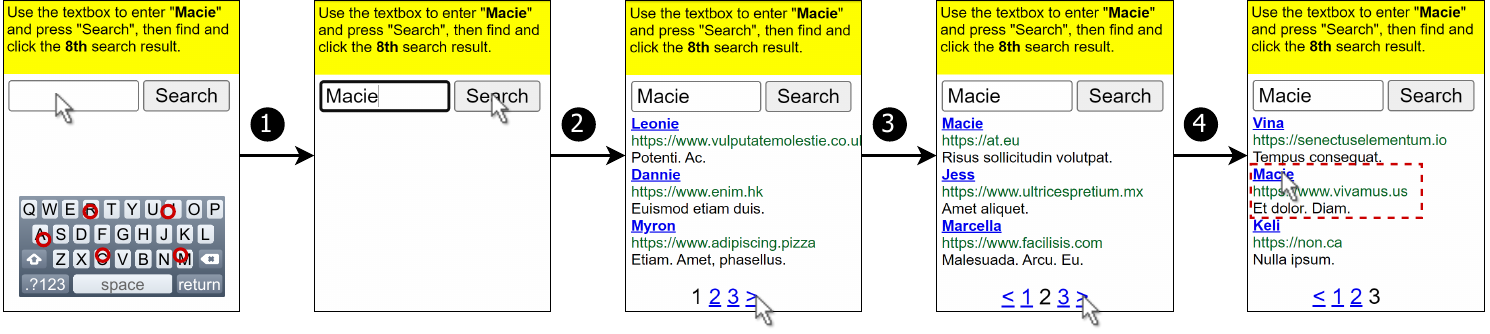}
\vspace{-12pt}
\caption{Step-by-step actions of an example \code{search-engine} task in the MiniWoB++ dataset 
\vspace{-12pt} 
~\cite{liu2018reinforcement}.}
\label{fig:motiv2}
\end{figure*}% 
An important step in web testing is generating a sequence of actions on the web pages from a natural language description of a task. 
%Such sequence forms the basis for a~testing environment to generate and run a test script.
Fig.~\ref{fig:motiv2} displays the task with the description: {\em``Use the textbox to enter `Macie' and press `Search', then find and click on the 8th search result.''} The tester is expected to follow this description and perform the described actions on the web pages under test. The tester uses a testing environment (Selenium or Katalon) to execute the application. First, the tester must type ``Macie'' into the textbox and click the ``Search'' button. They then need to click the ``$\geq$'' hyperlink twice before selecting the hyperlink associated with the entry ``Macie'' as shown in Step 4. 
%Figure~\ref{fig:motiv1} presents two tasks in each of which, a tester is provided with a natural-language description of the task. The tester is expected to follow this description and perform the described actions on the web pages under test. The tester uses a testing environment (Selenium or Katalon) to execute the application. In the first task, the picture shows the current page. (S)he follows the task description to enter the specified username and password into the textboxes and clicks ``Login'' button.
The process of the tester's performing a SoA (e.g., clicks, inputs) to interact with the environment (browser), ultimately changes the application's states (e.g., GUI, HTML, or screenshot).

%can be modeled as decision-making problems.

%, finding an email by a name and replying with specific text, and searching for the cheapest one-way flight from one place to another on a certain date.

\section{State-of-the-art Approaches}

%There are approaches designed to automate this manual process. Early methods introduced structured, domain-specific languages to help testers define actions and behaviors without delving into implementation (e.g., Gherkin, Cucumber~\cite{cucumber}).

\underline{First}, several LLM-based solutions have emerged to automate the manual process~\cite{gur2023real}. 
First, the {\bf UI agents for Web task execution} such as UI-TARS~\cite{qin2025uitarspioneeringautomatedgui}, SeeClick~\cite{cheng2024seeclick}, SeeAct~\cite{pmlr-v235-zheng24e}, CogAgent~\cite{hong2024cogagent}, and the models used in WebArena~\cite{zhou2023webarena} and WebLinx~\cite{lu2024weblinx}, focus on completing the task of "clicking on the 8th search result", i.e., the link with the word "Macie" (Fig.~\ref{fig:motiv2}). They {\em do not explicitly generate the executable actions that are needed for test script generation for Web applications}. For example, the WebArena benchmark would check only whether a model finally clicks on the "Macie" link. 

\underline{Second}, in contrast, UI-TARS~\cite{qin2025uitarspioneeringautomatedgui}, SeeClick~\cite{cheng2024seeclick}, and CogAgent~\cite{hong2024cogagent} generate the sequences of actions (SoAs) where an action is expressed as an operation on a screen area defined by {\bf absolute coordinates}, e.g., a "click" on [0,25,75,90] for the "Macie" link. {\em This is an unreliable locator approach for regression testing}, since even small UI modifications (e.g., shifting, resizing, or adding elements) can break coordinate-based references. It is not trivial to transform an action description with absolute coordinates into executable actions that are suitable for regression test script generation. 

\underline{Third}, another prominent line of approaches describe {\bf the actions or SoA in natural language}. In our example, at step 4, SeeAct~\cite{pmlr-v235-zheng24e} could describe the action as {\em "click on the word "Macie""}. However, there are two appearances of that word on the screen after step 4, which might make the model confused. In fact, it was reported that SeeAct suffers much lower accuracy in SoA generation when outputting executable actions~\cite{pmlr-v235-zheng24e}, which are crucial for Web test script generation.

% \vspace{-5pt}
\underline{Fourth}, several {\bf prompt-based approaches with demonstrations} such as MindAct~\cite{deng2024mind2web}, RCI~\cite{kim2024language}, AdaPlanner~\cite{sun2024adaplanner}, and Synapse~\cite{zheng2023synapse} leverage {\em few-shot or many-shot prompting} by providing exemplars to the LLMs. As shown in Fig.~\ref{fig:motiv2}, these demonstrations involve sequences like text entry and result navigation. However, they are limited by the heavy manual effort required for curation and the risk of overfitting to specific patterns.

% In our motivating example (Fig.~\ref{fig:motiv2}), a demonstration will contain a sequence of actions including typing a word into a search box, clicking on the "Search" button, clicking on a link for pagination, and clicking on a resulting link. However, these methods face two key limitations: (1) they demand substantial manual effort to create demonstrations on web UIs, and (2) they risk potential overfitting to the demonstrated patterns.

\underline{Fifth}, the prompt-based approaches without demonstrations leverage the {\bf self-reflection} capability of LLMs to identify and revise the first critical mistake after a failed attempt, and retry the task. However, {\em they operate in a reactive repairing strategy, where corrections are attempted only after an entire action trajectory has failed}.
%they operate with limited contextual awareness—it lacks information about the current page and feasible actions, and comprehensive execution history. 
Moreover, without maintaining a record of prior state-action pairs, for instance, at step 3 in Fig.~\ref{fig:motiv2}, the model might mistakenly click the hyperlink labeled “1,” returning to the previous screen. A heuristic based on textual similarity (e.g., matching the word “Macie”) fails, as multiple links share that label in steps 3 and 4. Relying on DOM indices is also unreliable, as not all search results are visible on a single page due to pagination.

%% file: sections/experience-memory.tex
\section{Long-term Experience Reinforcement} 
% \label{sec:experience}
% [Describe algorithm verbally]

% [Describe with example?]

%To enhance the central LLM's reasoning \circlednum[response]{2} performance of iterative agent planning, we incorporate Experience, which archives historical correct/incorrect sequences of actions and textual rules. 

%This experience reinforcement operates within a two-phased reflection strategy: the training phase and the evaluation phase. 

\algdef{SE}% flags used internally to indicate we're defining a new block statement
[CLASS]% new block type, not to be confused with loops or if-statements
{Class}% "\Struct{name}" will indicate the start of the struct declaration
{EndClass}% "\EndStruct" ends the block indent
[1]% There is one argument, which is the name of the data structure
{\textbf{class} \textsc{#1}}% typesetting of the start of a struct
{\textbf{end class}}% typesetting the end of the struct

%\Class{Experience}
%\State 
%\State $rules \gets []$

%============ Tien removed to save spaces for test generation ====
%\begin{algorithm}[t!]
%\footnotesize
%\caption{Experience Reinforcement Algorithm}
%\label{alg:algorithm2}
%\textbf{Input}: sequence actions memory $SAM$ of the current episode \\
%\begin{algorithmic}[1] %[1] enables line numbers
%    \State $correctSeq \gets []$, $inCorrectSeq \gets []$, $rules \gets []$
%    \Procedure{Update}{$SAM$}
%        \If{$SAM$ is correct}
%            \State $correctSeq.append(SAM)$
%        \Else
%            \State $inCorrectSeq.append(SAM)$
%            \State $rule \gets \textbf{LLM}(correctSeq, incorrectSeq, rules)$
%            \State $rules.append(rule)$
%        \EndIf
%        \State \textbf{return} self
%    \EndProcedure
%\end{algorithmic}
%\end{algorithm}
%=======================================================================

% \begin{comment}
% \begin{figure*}[htpb]
%     \centering
%     \includegraphics[width=1\textwidth]{figures/experience.pdf} 
%     \caption{An example of Experience update during training phase by {\tool} on MiniWoB++ task \textit{click-tab-2}.}
%     % \vspace{-2mm}
%     \label{fig:experience_visual}
% \end{figure*}
% \end{comment}

%Algorithm \ref{alg:algorithm2} presents the pseudo-code for our Experience Reinforcement algorithm. 

{\color{custom-blue}{
\textbf{Rule extraction and storage.} The Rule Extractor receives (i) the just-completed trajectory with its success/failure label and (ii) the full set of existing rules, and is prompted to extract a new textual rule that generalizes the observed behavior. Because all current rules are provided as part of this prompt, the LLM produces a rule that differs from those already stored, so the set is not duplicated and grows only when a genuinely new heuristic emerges. The extracted rules act as textual heuristics and, together with the correct SoAs serving as few-shot examples, populate the experience section of the main prompt (Fig.~9), which is updated dynamically during the training phase. We do not yet perform proactive consolidation or pruning of stored experience; for larger or longer-lived stores, consolidation of similar trajectories into canonical rules, utility- or age-based pruning, and periodic revalidation of stored rules against the current UI are natural extensions, which we leave to future work.}

\subsubsection*{\bf Training phase.} Initially, the experience is set to empty. During the training phase, the experience is updated once the task is successful. For update, the (in)correctly generated sequences of actions are saved as experience. For an incorrect sequence, we use Rule Extractor \circlednum[response]{4} to extract a textual rule from the (in)correct sequence, ensuring it differs from the current rules, and update it to the current set of rules. These rules act as textual heuristics, along with the correct SoA serving as few-shot examples embedded in the experience section of the main prompt (Fig.~\ref{prompt:experience_prompt}). This part of the main prompt is dynamically updated in the training phase. We also track the average of the correct SoA generated over a number of recent episodes in training (called {\em moving average}). 
Consider~Experience Reinforcement with task \code{click-tab-2}. The first episode initializes the task: “\textit{Switch between the tabs to find and click the link ‘et’}.” \tool uses Iterative LLM-based Agent Planning to generate and execute the action “\textit{click the link ‘et’},” completing the task. The correct sequence is then stored in the Experience (Algorithm \ref{alg:algorithm1}, lines 8–13). In the second episode, a similar task is initialized. The Experience guides the agent, but this time the generated sequence is incorrect. It is saved as an incorrect sequence, and Rule~Extractor creates the rules. In the third episode, with updated Experience (correct sequences+rules), it {\em switches tabs} before clicking the link, creating a correct sequence. This is added to Experience (Algorithm \ref{alg:algorithm1}, lines 14–18). Specifically, if the task remains unfinished at $t = MAX_t$ (the maximum number of steps), it is categorized as an incorrect execution (Algorithm \ref{alg:algorithm1}, lines 32–35).

\vspace{-5pt}
\subsubsection*{\bf Evaluation phase.} We select the optimal experience from training to be used as input for the evaluation phase. An optimal experience is the combination of the correct SoA and rules that exhibits an upward trend and stable pattern from the training, decided by the moving average. The iterative planning is the same, except for the fixed optimal experience $E_{optimal}$. Hence, with experience, the reasoning for the next action $a_{t+1}$ at the iteration $t$ become $a_{t+1} \sim \pi(. \mid o_t, \mathcal{A}(o_t), M_t, E_{optimal})$. 
% \begin{equation}
% a_{i+1} \sim \mathcal{R}_{LLM}(\cdot \mid \mathcal{O}_{1}^i,\mathcal{O}_{2}^i, \mathcal{O}_{3}^{opt})
% \end{equation}

%% file: sections/planning_appendix.tex
\section{Iterative Agent Planning}
\subsection{Implementation Details}
{\color{custom-blue}{
Algorithm~\ref{alg:algorithm1} displays the pseudo-code of our iterative LLM-based Agent Planning algorithm. It
takes as input a natural-language task description $task$ and a webpage $url$ for the process.  
% the Experience is initialized as empty (line 1).
First, the Sequence Actions Memory ($M_t$) is also initialized for the given task with an empty state-action history (line 2). 
The workflow begins by extracting either the DOM or screen images from the active webpage in the test environment (line 4). Next, an iterative planning phase is executed (lines 6–31), starting with the Content Extractor, which parses the webpage to derive the initial state and a set of feasible actions (line 6).

At line 7, the LLM is invoked to reason whether the current trajectory has achieved the goal ($success = true$) or an early stopping condition has been met ($stop = true$). If either case occurs, the algorithm logs the corresponding outcome in the Experience $E$—$(M_t, 1)$ for success or $(M_t, 0)$ for failure—during training mode, and then terminates the loop (lines 8–19). This process allows both successful and failed trajectories to contribute to improving the Experience for future planning.

If the task is still ongoing, the LLM is queried to infer the next action $a_t$ given the current observations $(\hat{s}_t, \mathcal{A}(o_t))$, the Sequence Actions Memory $M_t$, and the Experience $E$ (line 20). In cases where the generated action is duplicated (i.e., multiple identical candidates appear), the LLM is invoked again using the visual observation $o_t^{img}$—a screenshot of the current webpage—to select the correct element (lines 21–23). When the chosen action corresponds to an input operation, the LLM is prompted once more to generate the appropriate content for the target input field (lines 24–26). The completed action is then appended to the Sequence Actions Memory together with the current state (line 28).

Next, the Test Environment executes the selected action on the webpage to update the interface and retrieve the new state $s_t$ (lines 29–30). The loop counter $t$ is incremented, and the process continues until the task is completed, an early stop condition is reached, or the iteration limit $MAX_t$ is exceeded. If the maximum number of iterations is reached without completion, the trajectory is recorded as a failed episode $(M_t, 0)$ when in training mode (lines 32–35). Finally, the Experience $E$ is updated through the function $\phi_{LLM}(\tau_t, E)$ to refine future action planning (line 37).
}}

Let us take the iteration \#4 of Fig.~\ref{fig:motiv2} as an example. The HTML content of the webpage and its action transition arrow represent a state-action pair. At the end of iteration 3, $M_t$ currently stores the state-action pairs of the history including the ones at 1) the iteration $t$=1 (No search results displayed, input text: ``Macie''), 2) the iteration $t$=2 (No search results displayed, input text: ``Macie'', clicking ``Search''), and 3) the iteration $t$=3 (three links displayed: ``Leonie'', ``Dannie'', ``Myron'', pagination links, and clicking the ``$\leq$'' hyperlink). Then, an action $a_3$ is performed to transit the $3^{rd}$ page to the $4^{th}$. Next, we extract the current state $s_4$ and feasible actions $A_4$ (including the input search field, clicking Search button, clicking the ``Macie'', ``Jess'', and ``Marcella'' hyperlinks, clicking the ``$\leq$'' hyperlink, etc.). At the iteration $t=4$, the LLM reasons and decides the next action $a_4$ to be "clicking the ``$\geq$'' hyperlink" based on the observations of current webpage state $s_4$ and its corresponding actions $A_4$, the sequence actions memory $M_t$ up to iteration $t_3$, and the experience $E$. Then the procedure at the end of iteration $t=4$ continues and moves on to $t=5$ until the task is finished. Due to maintaining the contents/states of the steps, it does not stop and click at the ``Macie'' hyperlink at the 4th search result, and continues to ``Macie'' link in the next iteration. 
%Next, we will explain {\tool}'s key components.

\subsection{Task Description \& Content Extractor}

%goal given to the agent is stored in the Sequence Actions Memory 
%This task is defined by a user to be very explicit, leaving no ambiguity on the task.

The input task is described in natural language.
%(e.g., \textit{“Enter the username ‘Macie’ and the password ‘tsG’ into the text fields and press the ‘Login’ button”} as shown in Fig.~\ref{fig:motiv1}). 
This description is recorded in the Sequence Actions Memory (line 1, Algorithm~\ref{alg:algorithm1}). {\tool} utilizes Content Extractor \circlednum[response]{1} (lines 6, Algorithm~\ref{alg:algorithm1}) to extract information from the initial webpage and update the information after the agent performs actions (lines 28, Algorithm~\ref{alg:algorithm1}). The webpage is provided to the LLM either in DOM format or as screen images. This module has two functions: (1) feasible action extraction and (2) state extraction.

First, the LLM receives the DOM of the page and generates possible actions described as JSON objects. We extract only the visible and interactable HTML elements attached to specific user events using the DevTools API. We also extract relevant surrounding contexts. For example, the context for a button and a container could be its inner content, while an input field's context could be its label. If necessary, we traverse up the DOM tree to gather more information from parent nodes and bind accordingly. After that, each interactable element is mapped with its operation, resulting in a list of possible actions on the current webpage in a JSON file.
%(Fig.~\ref{jsonactions}).

Second, {\tool} extracts and represents the state of a webpage in one of two formats, based on the DOM size. For smaller DOMs, it generates a simplified HTML containing interactable elements and a mapping to the original HTML file. For large DOMs that exceed the LLM's context, it produces a text summary generated by the LLM from the webpage screenshot using vision capability (Fig.~\ref{listing:statenl}).

\begin{figure}[t]
\centering
\noindent % Use this command to remove  indentation before the lstlisting environment
% \lstset{style=textStyle}
\lstset{
    style=textStyle,
    breaklines=true,
    breakatwhitespace=false,    
    columns=fullflexible,
    linewidth=\columnwidth   
}
\begin{lstlisting}
The search results are displayed in a div with the id 'page-content'.
The first three search results are: 'Leonie', 'Dannie', and 'Myron' with additional links and descriptions.
There is also a pagination section with links to navigate through the pages.
\end{lstlisting}
\vspace{-14pt}
\caption{State in summarized natural language description format extracted from $3^{rd}$ webpage of Figure \ref{fig:motiv2}.}
\vspace{-12pt}
\label{listing:statenl}
\end{figure}

%An example of such a summary is illustrated in Fig.~\ref{listing:statenl}.

% Structure of the main prompt
\begin{figure}[t]
	\centering
	\lstset{
		numbers=left,
		numberstyle= \tiny,
		keywordstyle= \color{blue!70},
		commentstyle= \color{red!50!green!50!blue!50},
		frame=shadowbox,
		rulesepcolor= \color{red!20!green!20!blue!20} ,
		xleftmargin=1.5em,xrightmargin=0em, aboveskip=1em,
		framexleftmargin=1.5em,
                numbersep= 5pt,
		language=Json,
    basicstyle=\tiny	\ttfamily,
    numberstyle=\tiny	\ttfamily,
    emphstyle=\bfseries,
                % moredelim=**[is][\color{red}]{@}{@},
		% escapeinside= {(*@}{@*)}
	}

\begin{lstlisting}[style=pythonStyle]
ITERATIVE_ACTION_REASONING_PROMPT = '''You are a web assistant... You will complete the task by taking a series of steps. Each step is a description of the action you take and the specific item, entity, or element on the website that the action is applied.
(*@\bluetext{\{experience\}}@*)
# Here is the actual task.
(*@\bluetext{\{sequence\_actions\_memory\}}@*)
After completing the above steps, you reach a state: (*@\bluetext{\{state\}}@*) where the following feasible steps exist:
(*@\bluetext{\{feasible\_actions\}}@*)
POSSIBLE NEXT ACTION #1: (*@\bluetext{\{action\_1\}}@*) ...
Your job is to choose the most possible next steps to help you complete the task....
# The JSON response must strictly follow these rules:
    "chosen_action": ... (the index of the potential action that you choose)
    "action_description": ... (a string describing the action you choose)
    "reason": (describing why you choose the action)...
\end{lstlisting}
\vspace{-16pt}
\caption{Iterative Action Reasoning Prompt (Main Prompt).}
\label{prompt:iterative_prompt}
\vspace{-12pt}
\end{figure}

\begin{figure}
\begin{minipage}{0.48\textwidth}
\centering
	\lstset{
		numbers=left,
		numberstyle= \tiny,
		keywordstyle= \color{blue!70},
		commentstyle= \color{red!50!green!50!blue!50},
		frame=shadowbox,
		rulesepcolor= \color{red!20!green!20!blue!20} ,
		xleftmargin=1.5em,xrightmargin=0em, aboveskip=1em,
		framexleftmargin=1.5em,
                numbersep= 5pt,
		language=Json,
    basicstyle=\scriptsize\ttfamily,
    numberstyle=\scriptsize\ttfamily,
    emphstyle=\bfseries,
                % moredelim=**[is][\color{red}]{@}{@},
		% escapeinside= {(*@}{@*)}
	}

\begin{lstlisting}[style=pythonStyle]
SEQUENCE_ACTIONS_MEMORY_PROMPT = '''
You are visiting the website title: (*@\bluetext{\{title\}}@*)
You are asked to complete the following task: (*@\bluetext{\{task\}}@*)
You have completed the following steps: 
> STATE #1: (*@\bluetext{\{state\_1\}}@*)
> STEP #1: (*@\bluetext{\{action\_1\}}@*)
> STATE #2: (*@\bluetext{\{state\_2\}}@*)
> STEP #2: (*@\bluetext{\{action\_2\}}@*) ...
\end{lstlisting}
\vspace{-16pt}
\caption{Sequence Actions Memory Section of Main Prompt.}
\label{prompt:seq_mem_prompt}

\end{minipage}

\hfill
\begin{minipage}{0.48\textwidth}
\centering
	\lstset{
		numbers=left,
		numberstyle= \tiny,
		keywordstyle= \color{blue!70},
		commentstyle= \color{red!50!green!50!blue!50},
		frame=shadowbox,
		rulesepcolor= \color{red!20!green!20!blue!20} ,
		xleftmargin=1.5em,xrightmargin=0em, aboveskip=1em,
		framexleftmargin=1.5em,
                numbersep= 5pt,
		language=Json,
    basicstyle=\scriptsize\ttfamily,
    numberstyle=\scriptsize\ttfamily,
    emphstyle=\bfseries,
                % moredelim=**[is][\color{red}]{@}{@},
		% escapeinside= {(*@}{@*)}
	}
\begin{lstlisting}[style=pythonStyle]
EXPERIENCE_PROMPT = '''
# Here are the history of your trials
SUCCESS TRIAL #1: Task: (*@\bluetext{\{task\_1\}}@*)
STEP #1: (*@\bluetext{\{task\_1\_action\_1\}}@*)
STEP #2: (*@\bluetext{\{task\_1\_action\_2\}}@*)
...
SUCCESS TRIAL #8: Task: (*@\bluetext{\{task\_8\}}@*)
STEP #1: (*@\bluetext{\{task\_8\_action\_1\}}@*)
STEP #2: (*@\bluetext{\{task\_8\_action\_2\}}@*)
...
# Rules extracted from past attempts, use to evaluate your policy:
RULE #1: (*@\bluetext{\{rule\_1\}}@*)
RULE #2: (*@\bluetext{\{rule\_2\}}@*)...
\end{lstlisting}
\vspace{-16pt}
\caption{Experience Section of Main Prompt.}
\vspace{-12pt}
\label{prompt:experience_prompt}
\end{minipage}
\end{figure}

%% file: sections/dataset_appendix.tex
\section{Datasets}
This appendix complements Section~\ref{sec:eval} by presenting detailed descriptions and statistics of the datasets employed in our experiences.
The first, the Real-world dataset (Table~\ref{tab:realworld_data_stat_rq1}), was collected from a set of well-known websites including YouTube, LinkedIn, Facebook, Google, Amazon, StackOverflow, and Expedia. For each website $url$, we designed a set of task templates (e.g., \textit{“Subscribe to the \{\} channel”}, \textit{“Login to YouTube with username \{\} and password \{\}”}). Each template was then instantiated into specific tasks (e.g., \textit{“Subscribe to the `TEDx Talks' channel”}). The final dataset consists of 50 task instances per website across 7 websites, yielding a total of 350 task instances. 

The second dataset is from the {\bf MiniWoB++} benchmark~\cite{liu2018reinforcement}. Each environment or application is defined for a task, having a description (e.g., \textit{``Choose an item from a list''}), a set of utterances (e.g., \textit{``Bobine''}, \textit{``Betty''}) that were randomly chosen for each episode to form a complete natural language task instance (e.g., \textit{``Select Betty from the list..."}). We sampled 44 tasks from the dataset, each with 25 instances, yielding a total of 1,100 task instances.

{\color{my-blue}{The third dataset is {\bf OnlineMind2Web} \cite{xue2025onlineMind2Web}, a recent and actively maintained benchmark derived from the original Mind2Web dataset \cite{deng2024mind2web}. It contains 300 tasks collected from 136 real-world websites across domains such as clothing, food, housing, and transportation. The dataset classifies tasks into three difficulty levels: {\em easy} for those requiring up to 5 steps to complete (83 tasks), {\em medium} for those requiring 6–10 steps (143 tasks), and {\em hard} for those involving 11 steps or more (74 tasks). In our experiments, 15 tasks were excluded because four websites were inaccessible, resulting in a final set of 285 tasks from 132 websites used for evaluation.  
}}

%Table~\ref{tab:realworld_data_stat_rq1} reports for each web application the average number of actions per screen, the total number of screens, and the length of correct action sequences for tasks.

%The dataset presented in Table \ref{tab:realworld_data_stat_rq1} outlines real-world dataset statistics for a selection of 6 web applications, focusing on key metrics including: 1) Actions/Screens: the average number of feasible user interactions, accompanied by a standard deviation, 2) Screens: the average number of different screens or pages a user interacts with, and 3) ActionsSeq: the average length of correct sequence of actions.

%\begingroup
%\setlength{\tabcolsep}{6pt} % Default value: 6pt
%\renewcommand{\arraystretch}{1.2} % Default value: 1

\begin{table}[t]
\begin{minipage}{0.48\textwidth}
\footnotesize
\centering
\tabcolsep 3pt
\caption{\underline{Real-world} dataset statistics.}
\vspace{-6pt}
\label{tab:realworld_data_stat_rq1}
%\resizebox{\columnwidth}{!}{%
\begin{tabular}{|c|c|c|c|}
\hline
\textbf{Application} & \textbf{\#Actions/Screens} & \textbf{\#Screens} & \textbf{SoAs} \\ \hline
YouTube                  & $291\pm144$       & $2\pm1$         & $3\pm1$    \\
LinkedIn                 & $146\pm36$               & $3\pm1$         & $4\pm1$     \\
Facebook                 & $260\pm110$               & $3\pm1$         & $3\pm1$ \\
Google                   & $91\pm189$               & $3\pm1$          & $4\pm1$   \\
Amazon                   & $330\pm82$           & $2\pm1$         & $3\pm1$    \\
StackOverflow            & $296\pm151$          & $3\pm1$         & $5\pm5$          \\
Expedia            & $135\pm22$          & $4\pm1$         & $9\pm4$          \\ \hline
Total/Avg.               & 221               & 3               & 4          \\ \hline
\end{tabular}
\end{minipage}
\hfill
\begin{minipage}{0.48\textwidth}
\footnotesize
\caption{Baseline approaches}
\vspace{-6pt}
\label{tab1}
\centering
\tabcolsep 2.4pt
\begin{tabularx}{\columnwidth}{llccc}
\toprule
Approach &  Model & demos & Feedback\\
\midrule
WebNT5 & T5+Fine-tuning & \checkmark & - \\
HTML-T5-XL & T5+Fine-tuning & \checkmark & - \\
WebGUM & ViT+T5+Fine-tuning & \checkmark & - \\
\midrule
WGE & RL & \checkmark & - \\
CC-Net  & RL + SL & \checkmark & - \\
\midrule
RCI & LLM Prompting & \checkmark & \checkmark \\
AdaPlanner & LLM Prompting & \checkmark & \checkmark \\
Synapse & LLM Prompting & \checkmark & - \\
\midrule
Li et al. & LLM Prompting & - & \checkmark \\
SeeAct & LLM Prompting & - &  \checkmark \\
% \midrule
% Ours & Prompting & \checkmark & \checkmark & Generated & \checkmark \\
\bottomrule
\end{tabularx}
\label{tab:baseline_rq1}
\end{minipage}
\end{table}

%% file: sections/results_analyst.tex
\section{Result Analysis}
\subsection{\bf Analysis of the \underline{Real-world dataset}}
\label{sec:realworld_exp}

% {\tool} achieves an average of {\color{my-blue}{83.8\%}} \code{Exact-Match} and {\color{my-blue}{91.8\%}} \code{Prefix-Match} across all task instances. 

As seen in Table~\ref{tab:realworld_performance_rq1}, 
{\tool} outperforms Li {\em et al.} by {\color{my-blue}{56.8\%}}, with the gap being most significant in tasks that require navigating more than two screens and completing at least two actions. In this real-world dataset, websites are more dynamic than those in MiniWoB++, particularly when dealing with dropdown options (e.g., search tasks in \code{Google} or filling in locations to book flights on~\code{Expedia}, Table~\ref{tab:realworld_performance_rq1}). These dynamic features pose a challenge for the staged planning approach of Li {\em et al.}, which plans all actions on a screen at once. The reflection alone in Li {\em et al.} learns from the entire passing/failing sequences, thus, less flexible than its combination with ReAct and short-term memory as in {\tool}. Large state changes by an action often lead to the failure of subsequent actions in the plan. Moreover, Li {\em et al.} use the entire HTML for state representation, resulting in exceeding LLM context or selecting invalid actions.

In contrast, {\tool} performs well on the tasks in this dataset. These tasks involve state changes after an action, making the iterative planning in ReAct effective in adapting to changes. These tasks involve navigating up to four screens, performing up to five actions, processing large HTML files.

However, some tasks still pose greater challenges due to highly dynamic layouts and varying action sequences across runs. For instance, ordering a gift card may require navigating to the product detail page and filling out a ``Gift Card Details" form, whereas ordering a smartphone might only require clicking ``Add to Cart" after searching. These variations in layout/actions make planning with prior experience more difficult, leading to occasional inaccuracies. 

For \code{Expedia} tasks, it achieves the lowest \code{Exact-Match} of {\color{my-blue}{32\%}} and \code{Prefix-Match} score of {\color{my-blue}{64.4\%}}. It struggles to handle real-world date pickers to select departure and return dates.
\tool shows a performance with 80\% \code{Exact-Match} and 90\% \code{Prefix-Match} in \code{Stackoverflow} tasks. There is one task that requires filling a long form with constraints, resulting an incorrect actions.

{\color{my-blue}{
We also evaluated the best-performing variant of SeeAct~\cite{pmlr-v235-zheng24e}, \code{SeeAct$_{choice}$}, on the Real-world dataset. This model, which grounds actions using textual choices, was used to assess how a strong web-task model could be adapted to a new dataset. SeeAct achieved an \code{Exact-Match} rate of 72.4\% and a \code{Prefix-Match} rate of 83.8\%. It can handle most tasks on platforms such as \code{YouTube}, \code{Linked In}, \code{Google}, and \code{Amazon}. However, its performance drops to 66.7\% on \code{Facebook}, and it fails to complete tasks on \code{Expedia} due to difficulties handling real-world date pickers for selecting departure and return dates.
}}

\vspace{-9pt}
\subsection{\bf Analysis of the \underline{OnlineMind2Web}}
% \label{sec:rq1-online}

{\color{my-blue}{
Table \ref{tab:onlinemind2web_performance_rq1} shows the results obtained from running three approaches on the Online-Mind2Web dataset. {\tool} consistently outperforms the baselines across tasks in all difficulty levels. However, its performance varies by difficulty, with 81.3\% on easy tasks, 40.7\% on medium tasks, and 30.0\% on hard tasks. In this dataset, the websites are more complex than those in the Real-world dataset, particularly due to dropdown menus, calendar pickers, and similar components. Some UI elements are clickable while others are disabled, and task completion often depends on executing the correct sequence of clicks, further complicated by errors and formatting artifacts in the UIs. These dynamic features present significant challenges for the staged planning approaches of Li {\em et al.} and SeeAct. For Li {\em et al.}, using the DOM to represent the state of a webpage leads to token limitations in the LLM. In contrast, SeeAct suffers from incomplete descriptions of UI elements, which often result in suboptimal action choices.

For easy tasks, {\tool} has relative improvements of 2.03X over Li {\em et al.} and 27.4\% over SeeAct, with the substantial gains occurring in tasks that involve complex interface elements such as duplicate web elements, calendar pickers, and dropdown menus. For example, in the task {\em “Find Florida internship programs in the Mayo Clinic College of Medicine and Science,”} Li {\em et al.} was unable to reach the target page, whereas SeeAct failed as it attempted to resolve duplicate “Search” buttons.

For medium tasks, {\tool} achieves relative improvements of 2.75X over Li {\em et al.} and 33.9\% over SeeAct. These tasks require managing state changes triggered by user actions, where {\tool}'s iterative planning  is particularly effective for adapting to such dynamic interactions. For example, when a user enters a zip code into a search box, the system must wait for the resulting state change and the appearance of a dropdown list before selecting the appropriate item. In the task {\em “Browse pediatricians near zip code 90028 who specialize in Internal Medicine and have a rating of at least 4 stars,”} after the user inputs “90028” into the Location field, the system must wait for the state update and then correctly select {\em “Los Angeles, CA 90028”} from the dropdown.

For hard task, {\tool} achieves improvements of 5.3X over Li {\em et al.} and 3X over SeeAct. These tasks involve complex UI elements such as dropdowns and require multiple steps in the execution process. They also combine characteristics of both easy and medium tasks, as they mix relatively straightforward lookup operations with more intricate interactions that demand careful sequencing of actions. For instance, in the task {\em “Browse dermatologists within 10 miles of zip code 10019 and filter by only those who accept Blue Medicare Advantage,”} the agent must first enter the zip code “10019” into the Location field, wait for the interface to update, and then correctly select {\em “New York, NY 10019”} from the dropdown before applying the remaining filters. Satisfying the task requirements thus depends on completing the full sequence of steps. Such tasks often unfold as relatively long action chains, typically requiring three to four sub-actions per screen before progressing further.

Although {\tool} demonstrates substantial improvements over Li {\em et al.} and SeeAct, {\tool} still struggles with complex components such as datetime pickers, enter-to-submit search boxes, and highly intricate websites that may cause LLM context overflows.

\subsection{\bf Analysis of the \underline{MiniWoB++}}

\begin{table}[t]
\footnotesize
\centering
\caption{Action Sequence Generation on \underline{MiniWoB++}. Exact-match (\%), Correct runs (975 total).}
\label{tab:miniwob_analysic}
\vspace{6pt}
\begin{tabular}{lcc}
\toprule
\textbf{Approach} & \textbf{Exact-Match (\%)} & \textbf{\# Correct runs} \\ 
\midrule
WebNT5   & 59.1\% & 532 \\
HTML-T5  & 88.3\% & 861 \\
WebGUM   & 89.9\% & 877 \\ 
\midrule
WGE      & 64.3\% & 627 \\
CC-Net   & 93.8\% & 915 \\ 
\midrule
RCI      & 89.5\% & 926 \\
A.Plan.  & 88.0\% & 853 \\
Synapse  & 97.3\% & 958 \\ 
\midrule
Li (4o)  & 91.7\% & 949 \\
SeeAct   & 69.9\% & 682 \\
HxAgent  & 97.4\% & 950 \\
WALT     & 86.9\% & 847 \\ 
\bottomrule
\end{tabular}
\end{table}

% [Main table here - todo for today]
% Another version of main table
%\begingroup

% \begin{table}[t]
% \footnotesize
% \tabcolsep 2.2pt
% \centering
% \caption{{\color{my-blue}{Action Sequence Generation on \underline{MiniWoB++} (RQ1). Exact-match (\%), Correct runs (975 total).}}}
% \label{tab:miniwob_rq1}
% \vspace{-6pt}
% \begin{tabular}{|l||cc|}
% \hline
% \multicolumn{1}{|c||}{\multirow{2}{*}{\textbf{Approach}}} & \multicolumn{2}{c|}{\textbf{Metrics}} \\ \cline{2-3} 
% \multicolumn{1}{|c||}{} & \multicolumn{1}{c|}{\textbf{Exact-Match (\%)}} & \textbf{\# Correct runs} \\ \hline \hline
% WebNT5      & \multicolumn{1}{c|}{59.1\%} & 532 \\
% HTML-T5     & \multicolumn{1}{c|}{88.3\%} & 861 \\
% WebGUM      & \multicolumn{1}{c|}{89.9\%} & 877 \\ \hline
% WGE         & \multicolumn{1}{c|}{64.3\%} & 627 \\
% CC-Net      & \multicolumn{1}{c|}{93.8\%} & 915 \\ \hline
% RCI         & \multicolumn{1}{c|}{89.5\%} & 926 \\
% A.Plan.     & \multicolumn{1}{c|}{88.0\%} & 853 \\
% Synapse     & \multicolumn{1}{c|}{97.3\%} & 958 \\ \hline
% Li (4o)     & \multicolumn{1}{c|}{91.7\%} & 949 \\
% SeeAct      & \multicolumn{1}{c|}{69.9\%} & 682 \\
% HxAgent     & \multicolumn{1}{c|}{97.4\%} & 950 \\ \hline
% \end{tabular}

% \end{table}

As seen in Table \ref{tab:miniwob_rq1}, which reports 975 task instances from 39 tasks, {\tool} achieves a {\color{my-blue}{97.4\%}} average \code{Exact-Match} across the shared set of tasks among the baselines. This set of tasks varies in screens and actions, with screens ranging from 1 to 6, and a maximum of 6 actions needed to complete the task. Tasks with fewer screens and actions tend to have a higher average \code{Exact-Match}, with 97\% to 100\% reported for tasks with fewer than 3 screens and actions. This percentage drops marginally to 93\% for tasks containing more than 3 screens and actions, with a maximum of 4 screens and 9 actions. With \code{Prefix-Match} of 99\%, in many cases where \tool does not fully succeed in a task, the generated sequences contain correct prefixes of action sequences, reducing correction effort.

Although \tool achieves comparable \code{Exact-Match} with the best baselines in each category, note that it {\em eliminates the need for human demonstrations} (see Table \ref{tab:miniwob_analysic}). Specifically, it matches the performance of WebGUM, despite WebGUM utilizing 401K demonstrations with web screenshots to jointly fine-tune the vision encoder ViT and T5. \tool performed better than CC-Net (BC + RL) in 10 task instances, with CC-Net employing 2.4M human's demonstrations.
%collected from 77 human participants for behavior cloning. 
Synapse, with perfect results, relies heavily on 154 quality hand-crafted exemplars, some of which further contain task-specific filter prompts to convert raw HTML states into clean observations.
Their solution may result in overfitting MiniWoB++, as they have a high number of few-shot exemplars and the reducing of the difficulty of tasks requiring exploration in multiple states with customized and obvious~observation.

Among the approaches that do not rely on human demonstrations, {\tool} performs slightly better than Li {\em et al.} (97\% vs. 94\%) and substantially outperforms SeeAct by 20\% in \code{Exact-Match} accuracy. We further stratified results on 125 challenging task instances drawn from five MiniWoB++ tasks (Table~\ref{tab:additional_miniwob_RQ1}). These tasks are considered challenging because they require navigation across an average of five screens and the execution of approximately seven actions to complete. As seen in Table~\ref{tab:additional_miniwob_RQ1}, \tool shows the effectiveness {\color{my-blue}{({\bf 81.6\% \code{Exact-Match}})}} in more challenging tasks over Li {\em et al.} and SeeAct, which have only 10\% and 8\% \code{Exact-Match} in these complex actions.
% which grounds actions on descriptions of possible choices

{\color{my-blue}{
As reported in Table~\ref{tab:miniwob_rq1}, SeeAct achieves an \code{Exact-Match} rate of 69.9\%, which underperforms Li {\em et al.} On the 125 most challenging tasks, its \code{Exact-Match} rate drops to 8\%, with successful completion limited to the \code{flight.AA} tasks (Table~\ref{tab:additional_miniwob_RQ1}). Our further analysis reveals that web components, such as datetime pickers and dropdowns, remain significant obstacles for SeeAct. This is also true for other non-descriptive elements like buttons with only icons, which require action-oriented descriptions to be processed effectively. In brief, {\em {\tool} achieves higher accuracies on the challenging tasks, while the baselines struggle with such intricate actions}.
}}

\subsection{\bf Accuracy by Task Complexity}
We evaluate the performance of {\tool} with respect to task complexity, measured by the length of action sequences and the number of actions on webpages.

{\bf Next-step prediction accuracy by action sequence length}. 
In Fig.~\ref{fig:step-acc}, \tool achieves 100\% accuracy on the first step for all tasks in the MiniWoB++ and Real-world datasets. However, as a task requires more action steps, accuracy slightly decreases.
(For step 7, a slight increase is observed due to the smaller number of tasks reaching this stage, resulting in a relatively higher value). For two tasks reaching Step 11, it did not predict the correct action, despite the successes of the previous 10 steps. A similar trend is observed in the OnlineMind2Web dataset, where \code{Exact-Match} drops from 81.3\% on easy tasks to 40.7\% on medium tasks and 30.0\% on hard tasks.  

% \begin{figure}[t!]
%     \centering
%     \begin{subfigure}[t]{0.45\textwidth}
%         \centering
%         \includegraphics[width=\textwidth]{figures/step_accuracy.pdf}
%         \vspace{-15pt}
%         \caption{Next-step Prediction.}
%         \label{fig:step-acc}
%     \end{subfigure}%
%     \hspace{1pt}% Space between image A and B
%     \begin{subfigure}[t]{0.45\textwidth}
%         \centering
%         \includegraphics[width=\textwidth]{figures/action_screens_accuracy.pdf}
%         \vspace{-15pt}
%         \caption{Accuracy by Actions/Screens.}
%         \label{fig:action-screen-acc}
%     \end{subfigure}
%      \vspace{-12pt}
%     \caption{Performance by Task Complexity (RQ1)}
%     % \vspace{-12pt}
%     \label{fig:experience_rq2}
% \end{figure}
% \vspace{2pt}
\noindent \emph{{\bf Exact-Match sequence accuracy by number of actions per screen}}. 
The action space may contain a large number of selectable actions per screen. As shown in Fig.~\ref{fig:action-screen-acc}, {\tool} maintains high \code{Exact-Match} accuracy even when the action space is large, with up to 500 possible actions per screen.

There are eight outliers (shown in orange): two tasks with sequence lengths of up to 11 steps, where \code{Exact-Match} falls below 70\%. For tasks involving dynamic content, such as confirmation popups in \code{linkedin-3} and story posts in \code{facebook-5}, {\tool} performs reasonably well but with limited consistency. Tasks requiring the selection of hidden options (\code{flight-Alaska-auto}) or actions that span multiple UI elements (e.g., date-pickers in \code{expedia-3} and \code{expedia-5}) achieve only 30\% \code{Exact-Match}. Similarly, for tasks involving complex forms (e.g., \code{stackoverflow-3}), the system generated invalid data during form filling.

\subsection{\bf Cost Analysis }
\label{sec:CostAnalsysis}

\begin{figure*}[htpb]
    \centering
\includegraphics[width=0.98\textwidth]{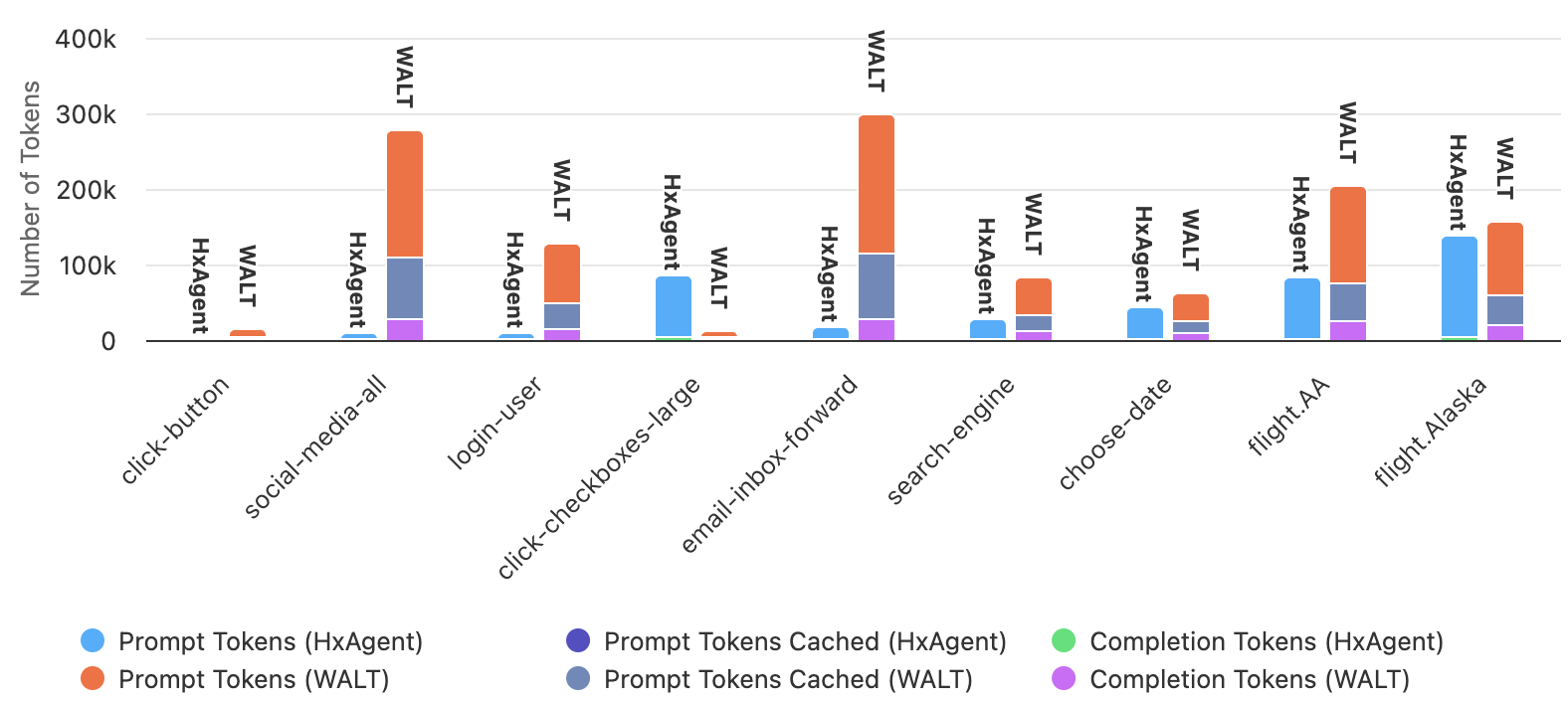} 
    \vspace{-6pt}
    \caption{Token Consumption Comparison by Dataset and Tool.}
    \vspace{-12pt} 
    \label{fig:token_consumption}
\end{figure*}
We conduct a cost analysis on representative MiniWob++ tasks, ranging from simple tasks such as `click-button` and `login-user` to more complex ones including `flight.AA` and `flight.Alaska`. Figure~\ref{fig:token_consumption} shows the total token consumption of each method.

The results show that {\tool} achieves lower and more stable token usage than the baselines, especially on tasks with complex interfaces or long interaction sequences. In contrast, WALT consistently consumes substantially more tokens, particularly on `social-media-all`, `email-inbox-forward`, and `flight.AA`, indicating higher inference costs for handling complex UI interactions.

Tasks involving dynamic forms, date pickers, and multi-step navigation generally increase token consumption across all methods. Nevertheless, {\tool} remains more cost-efficient while achieving competitive or superior performance. Compared with WALT, \tool uses only about one-third of WALT’s total tokens, while its output tokens account for just 8.9\% of WALT’s output token count.

\begin{figure}
\begin{minipage}{0.48\textwidth}
\centering
    \includegraphics[width=\textwidth]{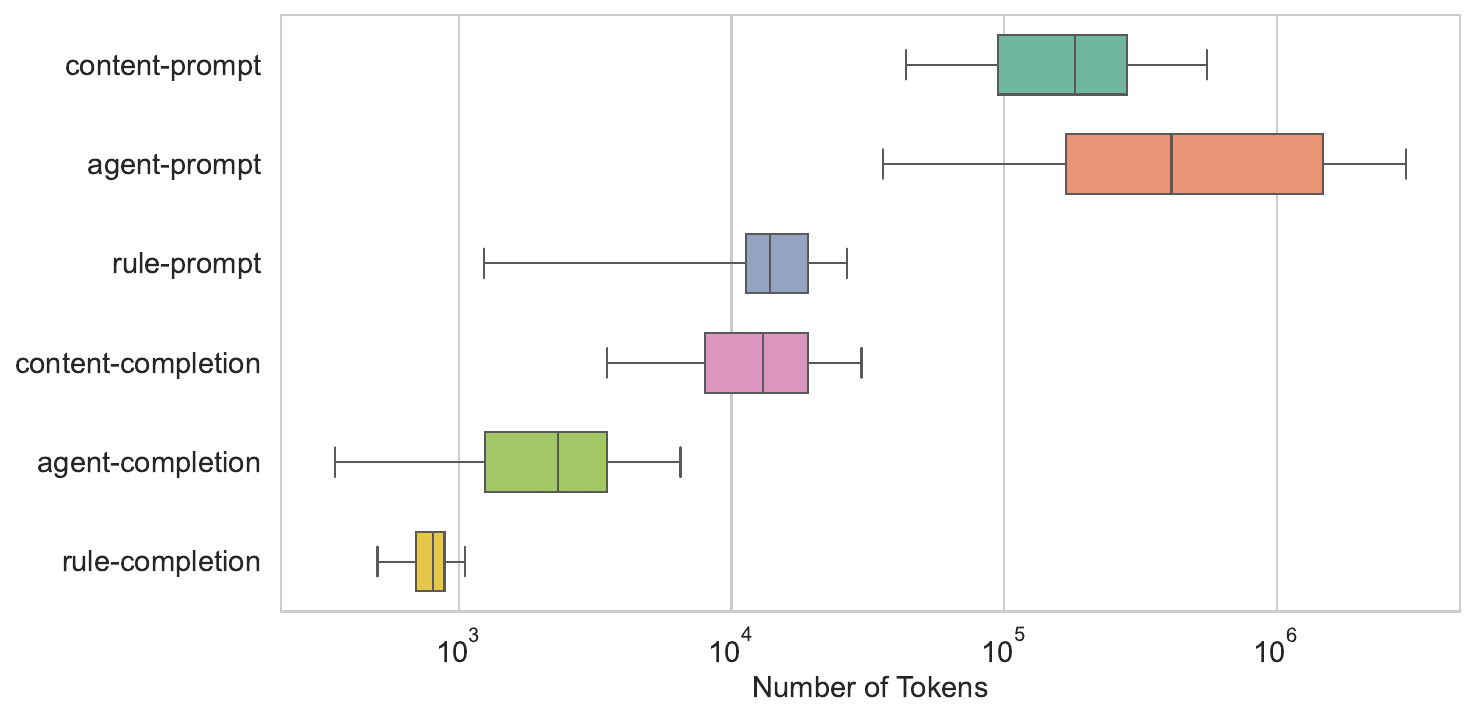} 
    \vspace{-18pt}
    \caption{\#tokens for each component's input/output.}
    \label{fig:token_main}
\end{minipage}
% \hfill
\begin{minipage}{0.48\textwidth}
\centering
    \includegraphics[width=2.7in]{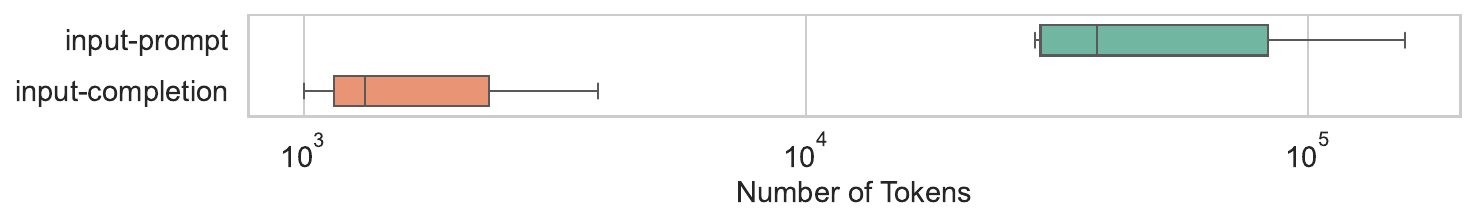} 
    \vspace{-9pt}
    \caption{\#tokens for input generator (log scale).}
    \label{fig:token_input}
\end{minipage}
\end{figure}

Figures \ref{fig:token_main} and \ref{fig:token_input} provide a detailed breakdown of token consumption across different components. Figure \ref{fig:token_main} shows that the agent module consumes the largest number of tokens, particularly in the prompt stage, indicating that reasoning and action planning dominate the inference cost. In contrast, the rule-completion component requires relatively few tokens.

Figure \ref{fig:token_input} further shows that, within the input generator, prompt tokens account for most of the cost, while completion tokens remain comparatively small. Overall, these results suggest that contextual grounding and webpage representation are the primary sources of token consumption in the framework.